\documentclass[twocolumn]{IEEEtran}
\usepackage{amsmath}
\usepackage{amssymb}
\usepackage{amsfonts}
\usepackage{slashbox}
\usepackage{graphicx}
\usepackage{cite, amsmath, amsfonts, amssymb, psfrag, epsfig,tikz}
\usetikzlibrary{shapes,shadows,arrows}
\newtheorem{proposition}{{Proposition}}

\newtheorem{theorem}{{Theorem}}
\newtheorem{lemma}{{Lemma}}

\DeclareMathAlphabet{\mathpzc}{OT1}{pzc}{m}{it}

\begin{document}

\tikzstyle{line} = [draw, -stealth,semithick]
\tikzstyle{block} = [draw, rectangle, text width=7em, text centered, minimum height=12mm, node distance=8em,semithick]
\tikzstyle{chblock} = [draw, rectangle, text width=4em, text centered, minimum height=40mm, node distance=8em,semithick]

\title{When is Noisy State Information at the Encoder as Useless as No Information or as Good as Noise-Free State?}


\author{Rui Xu, Jun Chen, Tsachy Weissman, and Jian-Kang Zhang
\thanks{Rui Xu, Jun Chen, and Jian-Kang Zhang are with the Department
of Electrical and Computer Engineering, McMaster University,
Hamilton, ON L8S 4K1, Canada  (email: xur27@mcmaster.ca \{junchen,jkzhang\}@ece.mcmaster.ca).}
\thanks{Tsachy Weissman is with the Department of Electrical Engineering, Stanford University, Stanford, CA 94305, USA (email: tsachy@stanford.edu).}}

\maketitle

\begin{abstract}
For any binary-input channel with perfect state information at the decoder, if the mutual information between the noisy state observation at the encoder and the true channel state is below a positive threshold determined solely by the state distribution, then the capacity is the same as that with no encoder side information. A complementary phenomenon is revealed for the generalized probing capacity. Extensions beyond binary-input channels are developed.



\end{abstract}


\section{Introduction}\label{sec:introduction}

\begin{figure*}[tb]
\begin{centering}
\includegraphics[width=13cm]{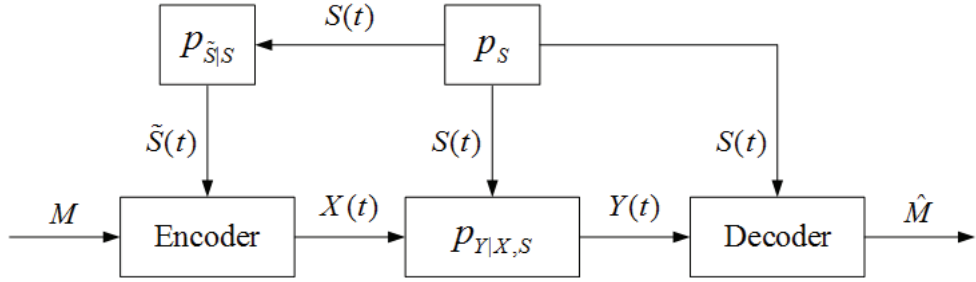}
\caption{Channel model.\label{fig:model}}
\end{centering}
\end{figure*}

Consider a memoryless channel $p_{Y|X,S}$ with input $X$, output $Y$, and state $S$. We assume that the channel state $S$, distributed according to $p_S$, is provided to the decoder, and a noisy state observation $\tilde{S}$, generated by $S$ through side channel $p_{\tilde{S}|S}$, is available causally at the encoder. Here $X$, $Y$, $S$, and $\tilde{S}$ are defined over finite alphabets $\mathcal{X}$, $\mathcal{Y}$, $\mathcal{S}$, and $\tilde{\mathcal{S}}$, respectively.
In this setting (see Fig. \ref{fig:model}), Shannon's remarkable result \cite{Shannon58} (see also \cite[Eq. (3)]{CS99} and \cite[Th. 7.2]{EGK11}) implies that the channel capacity is given by
\begin{align}
C(p_{Y|X,S},p_S,p_{\tilde{S}|S})\triangleq\max\limits_{p_U} I(U;Y|S).\label{eq:noisystate}
\end{align}
The auxiliary random variable $U$ is defined over alphabet $\mathcal{U}$ with $|\mathcal{U}|=|\mathcal{X}|^{|\tilde{\mathcal{S}}|}$, 
whose joint distribution with $(X,Y,S,\tilde{S})$ factors as
\begin{align}
&p_{U,X,Y,S,\tilde{S}}(u,x,y,s,\tilde{s})\nonumber\\
&= p_U(u)p_S(s)p_{\tilde{S}|S}(\tilde{s}|s)\mathbb{I}(x=\psi(u,\tilde{s}))p_{Y|X,S}(y|x,s),\nonumber\\
&\hspace{1in}u\in\mathcal{U}, x\in\mathcal{X}, y\in\mathcal{Y}, s\in\mathcal{S}, \tilde{s}\in\tilde{S},\label{eq:jointdistribution}
\end{align}
where $\mathbb{I}(\cdot)$ is the indicator function, and $\psi(u,\cdot)$, $u\in\mathcal{U}$, are $|\mathcal{X}|^{|\tilde{\mathcal{S}}|}$ different mappings from $\tilde{\mathcal{S}}$ to $\mathcal{X}$.
Without loss of generality, we set $\mathcal{X}=\{0,1,\cdots,|\mathcal{X}|-1\}$, $\mathcal{S}=\{0,1\cdots,|\mathcal{S}|-1\}$,   $\mathcal{U}=\{0,1,\cdots,|\mathcal{X}|^{|\tilde{\mathcal{S}}|}-1\}$, and order the mappings $\psi(u,\cdot)$, $u\in\mathcal{U}$, in such a way that the first $|\mathcal{X}|$ mappings\footnote{These are the mappings that ignore the encoder side information.} are
\begin{align}
\psi(u,\cdot)\equiv u,\quad u\in\mathcal{X};\label{eq:orderpsi}
\end{align}
moreover, we assume that $\rho\triangleq\min_{s\in\mathcal{S}}p_S(s)>0$. The capacity formula (\ref{eq:noisystate}) can be simplified in the following two special cases.
Specifically, when there is no encoder side information, the channel capacity  reduces to \cite[Eq. (7.2)]{EGK11}
\begin{align}
\underline{C}(p_{Y|X,S},p_S)\triangleq\max\limits_{p_X}I(X;Y|S),\label{eq:nostate}
\end{align}
where $p_{X,Y,S}(x,y,s)= p_{X}(x)p_S(s)p_{Y|X,S}(y|x,s)$; on the other hand, when perfect state information is available at the encoder (as well as the decoder), the channel capacity  becomes \cite[Eq. (7.3)]{EGK11}
\begin{align}
\overline{C}(p_{Y|X,S},p_S)\triangleq\max\limits_{p_{X|S}}I(X;Y|S),\label{eq:perfectstate}
\end{align}
where $p_{X,Y,S}(x,y,s)= p_S(s)p_{X|S}(x|s)p_{Y|X,S}(y|x,s)$.

For comparison, consider the following similarly defined quantity
\begin{align*}
C'(p_{Y|X,S},p_S,p_{\tilde{S}|S})\triangleq\max\limits_{p_U}I(X;Y|S),
\end{align*}
where the joint distribution of $(U,X,Y,S,\tilde{S})$ is also given by (\ref{eq:jointdistribution}).
We shall refer to $C'(p_{Y|X,S},p_S,p_{\tilde{S}|S})$ as the generalized probing capacity. By the functional representation lemma \cite[p. 626]{EGK11} (see also \cite[Lemma 1]{WCZCP11}),  $C'(p_{Y|X,S},p_S,p_{\tilde{S}|S})$ can be defined equivalently as
\begin{align*}
C'(p_{Y|X,S},p_S,p_{\tilde{S}|S})\triangleq\max\limits_{p_{X|\tilde{S}}}I(X;Y|S),
\end{align*}
where
\begin{align*}
&p_{X,Y,S,\tilde{S}}(x,y,s,\tilde{s})\\
&= p_S(s)p_{\tilde{S}|S}(\tilde{s}|s)p_{X|\tilde{S}}(x|\tilde{s})p_{Y|X,S}(y|x,s),\\
&\hspace{1in}x\in\mathcal{X}, y\in\mathcal{Y}, s\in\mathcal{S}, \tilde{s}\in\tilde{\mathcal{S}}.
\end{align*}
Clearly,
\begin{align}
\underline{C}(p_{Y|X,S},p_S)&\leq C(p_{Y|X,S},p_S,p_{\tilde{S}|S})\nonumber\\
&\leq C'(p_{Y|X,S},p_S,p_{\tilde{S}|S})\nonumber\\
&\leq \overline{C}(p_{Y|X,S},p_S).\label{eq:comb1}
\end{align}
Moreover, we have
\begin{align}
C(p_{Y|X,S},p_S,p_{\tilde{S}|S})&= C'(p_{Y|X,S},p_S,p_{\tilde{S}|S})\nonumber\\
&=\underline{C}(p_{Y|X,S},p_S)\label{eq:endpoint1}
\end{align}
if $S$ and $\tilde{S}$ are independent (i.e., $I(S;\tilde{S})=0$), and
\begin{align}
C(p_{Y|X,S},p_S,p_{\tilde{S}|S})&= C'(p_{Y|X,S},p_S,p_{\tilde{S}|S})\nonumber\\
&=\overline{C}(p_{Y|X,S},p_S)\label{eq:endpoint2}
\end{align}
if $S$ is a deterministic function of $\tilde{S}$ (i.e., $H(S|\tilde{S})=0$).

\begin{figure*}[tb]
\begin{centering}
\includegraphics[width=8cm]{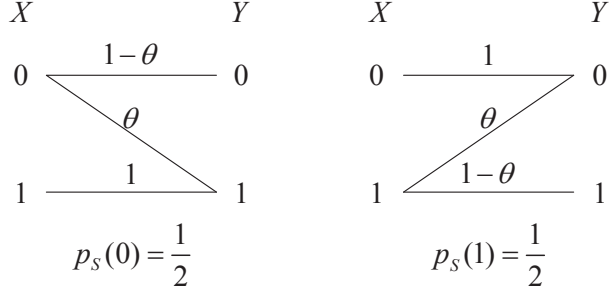}
\caption{Illustration of $p_{Y|X,S}$ and $p_S$ given by (\ref{eq:para1}) and (\ref{eq:para2}), respectively.\label{fig:SZchannel}}
\end{centering}
\end{figure*}


\begin{figure*}[tb]
\begin{centering}
\includegraphics[width=13cm]{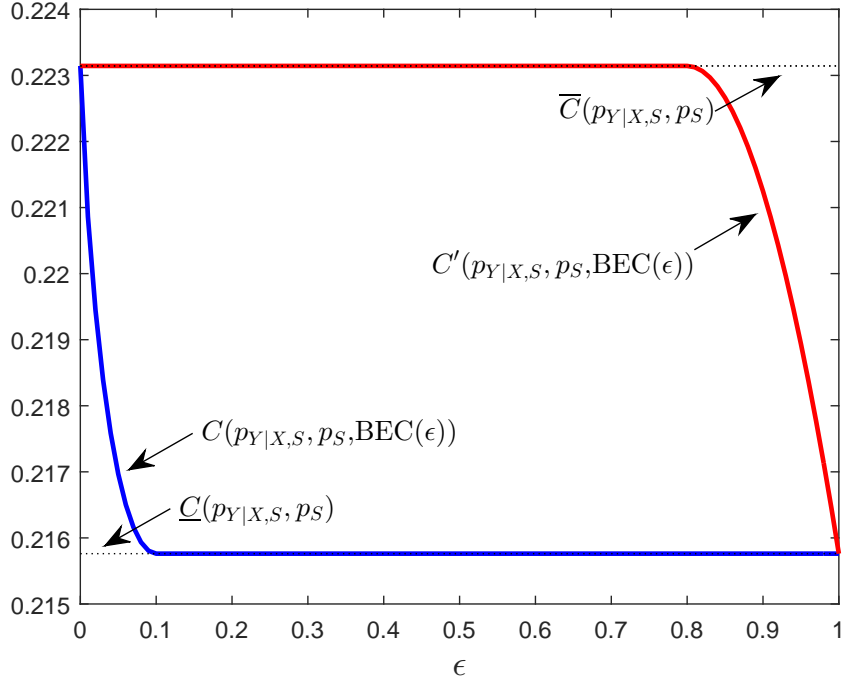}
\caption{Plots of $C(p_{Y|X,S},p_S,\mbox{BEC}(\epsilon))$ and $C'(p_{Y|X,S},p_S,\mbox{BEC}(\epsilon))$ against $\epsilon$ for $\epsilon\in[0,1]$, where $p_{Y|X,S}$ and $p_S$ are given by (\ref{eq:para1})  with $\theta=\frac{1}{2}$ and (\ref{eq:para2}), respectively. \label{fig:plot1}}
\end{centering}
\end{figure*}
\begin{figure*}[tb]
\begin{centering}
\includegraphics[width=13cm]{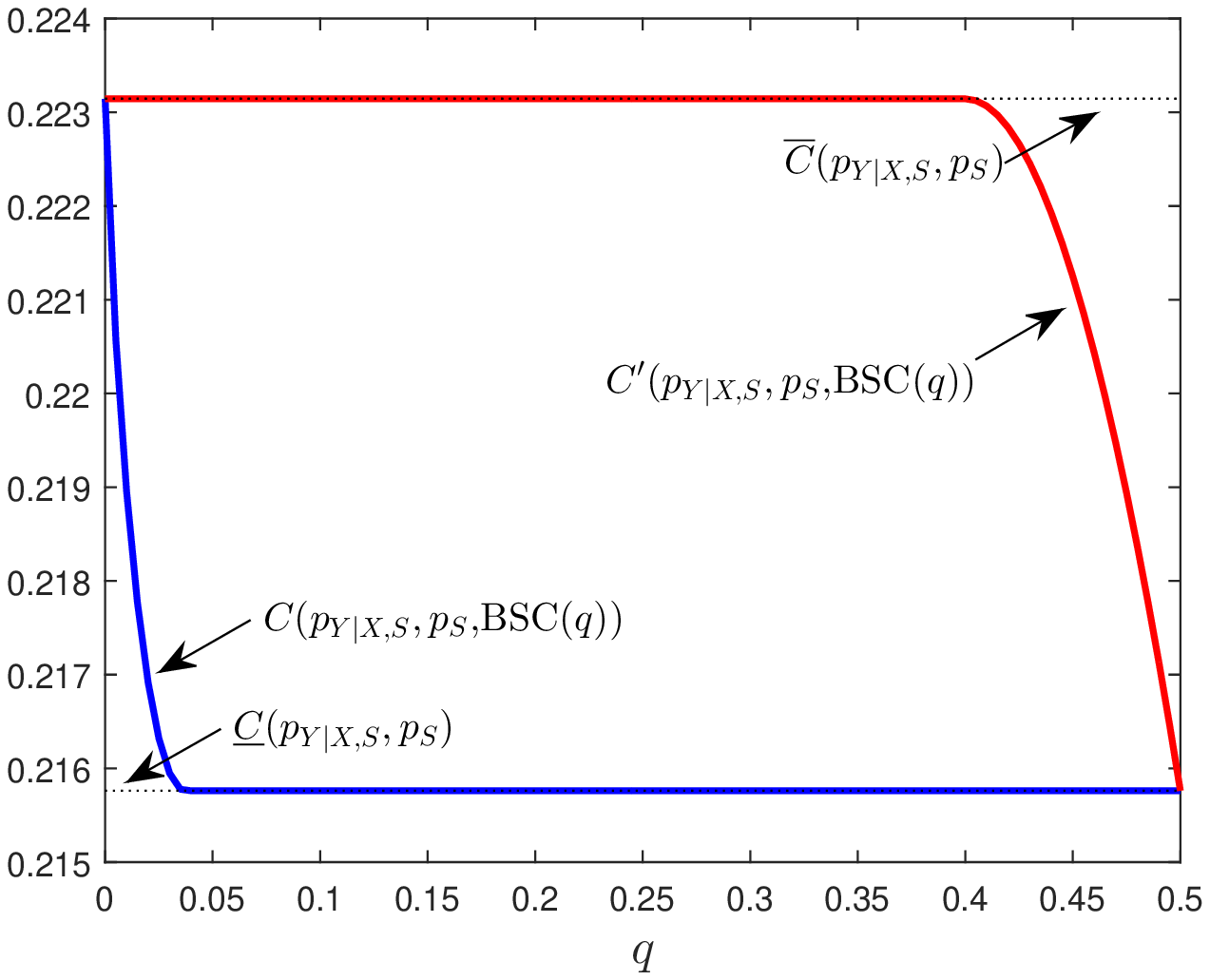}
\caption{Plots of $C(p_{Y|X,S},p_S,\mbox{BSC}(q))$ and $C'(p_{Y|X,S},p_S,\mbox{BSC}(q))$ against $q$ for $q\in[0,\frac{1}{2}]$, where $p_{Y|X,S}$ and $p_S$ are given by (\ref{eq:para1}) with $\theta=\frac{1}{2}$ and (\ref{eq:para2}), respectively.\label{fig:plot2}}
\end{centering}
\end{figure*}

To elucidate the operational meaning of $C'(p_{Y|X,S},p_S,p_{\tilde{S}|S})$ and its connection with $C(p_{Y|X,S},p_S,p_{\tilde{S}|S})$, it is instructive to consider the special case where  $p_{\tilde{S}|S}$ is a binary erasure channel with erasure probability $\epsilon$ (denoted by $\mbox{BEC}(\epsilon)$), which corresponds to the probing channel setup studied in \cite{APW11}. The probing channel model is essentially the same as the one in Fig. \ref{fig:model} except that, in Fig. \ref{fig:model}, the encoder (which, with high probability, observes approximately $n\epsilon$ state symbols out of the whole state sequence of length $n$ when $n$ is large enough) has no control of the exact positions of these $n\epsilon$ symbols whereas, in the probing channel model, the encoder has the freedom to specify the positions of these $n\epsilon$ symbols according to the message to be sent. It is shown in \cite{APW11} that this additional freedom increases the achievable rate from $C(p_{Y|X,S},p_S,\mbox{BEC}(\epsilon))$ to $C'(p_{Y|X,S},p_S,\mbox{BEC}(\epsilon))$.
Now consider an example (see also Fig. \ref{fig:SZchannel}) where 
\begin{align}
&p_{Y|X,S}(y|x,s)=\left\{
                   \begin{array}{ll}
                     1-\theta, & (x,y,s)=(0,0,0)\mbox{ or }(1,1,1), \\
                     \theta, & (x,y,s)=(0,1,0)\mbox{ or }(1,0,1), \\
                     0, & (x,y,s)=(1,0,0)\mbox{ or }(0,1,1),\\
                     1, & (x,y,s)=(1,1,0)\mbox{ or }(0,0,1),
                   \end{array}
                 \right.\label{eq:para1}\\
&p_S(0)=p_S(1)=\frac{1}{2}.\label{eq:para2}
\end{align}
For this example, it can be verified that
\begin{align*}
&\underline{C}(p_{Y|X,S},p_S)\\
&=\left\{
                                                    \begin{array}{ll}
                                                      \log 2, & \theta=0, \\
                                                      \frac{1}{2}\Bigg((1-\theta)\log 2+\log\frac{2}{1+\theta}+\theta\log\frac{2\theta}{1+\theta}\Bigg), & \theta\in(0,1), \\
                                                      0, & \theta=1,
                                                    \end{array}
                                                  \right.\\
&\overline{C}(p_{Y|X,S},p_S)=\left\{
                                                    \begin{array}{ll}
                                                      \log2, & \theta=0, \\
                                                      \log\Bigg(1+(1-\theta)\theta^{\frac{\theta}{1-\theta}}\Bigg), & \theta\in(0,1), \\
                                                      0, & \theta=1.
                                                    \end{array}
                                                  \right.
\end{align*}
Note that $\overline{C}(p_{Y|X,S},p_S)$ is strictly greater than $\underline{C}(p_{Y|X,S},p_S)$ unless $\theta=0$ or $\theta=1$.
It follows by (\ref{eq:endpoint1}) and (\ref{eq:endpoint2}) that
\begin{align*}
\left.C(p_{Y|X,S},p_S,\mbox{BEC}(\epsilon))\right|_{\epsilon=1}&= \left.C'(p_{Y|X,S},p_S,\mbox{BEC}(\epsilon))\right|_{\epsilon=1}\\
&=\underline{C}(p_{Y|X,S},p_S),\\
\left.C(p_{Y|X,S},p_S,\mbox{BEC}(\epsilon))\right|_{\epsilon=0}&= \left.C'(p_{Y|X,S},p_S,\mbox{BEC}(\epsilon))\right|_{\epsilon=0}\\
&=\overline{C}(p_{Y|X,S},p_S).
\end{align*}
To gain a better understanding, we plot $C(p_{Y|X,S},p_S,\mbox{BEC}(\epsilon))$ and $C'(p_{Y|X,S},p_S,\mbox{BEC}(\epsilon))$ against $\epsilon$ for $\epsilon\in[0,1]$ in Fig. \ref{fig:plot1}.
It turns out that, somewhat  counterintuitively,  $C(p_{Y|X,S},p_S,\mbox{BEC}(\epsilon))$ coincides with $\underline{C}(p_{Y|X,S},p_S)$ way before $\epsilon$ reaches 1. That is to say, when $\epsilon$ is above a certain threshold strictly less than 1, the noisy state observation $\tilde{S}$ 
is useless and can be ignored (as far as the channel capacity is concerned). On the the hand, it can be seen that
$C'(p_{Y|X,S},p_S,\mbox{BEC}(\epsilon))$ is equal to $\overline{C}(p_{Y|X,S},p_S)$ for a large range of $\epsilon$ strictly greater than 0. Hence, in terms of the probing capacity, the noisy state observation can be as good as the perfect one.
As shown in Fig. \ref{fig:plot2}, the same phenomena arise if we choose $p_{\tilde{S}|S}$ to be a binary symmetric channel with crossover probability $q$ (denoted by $\mbox{BSC}(q)$).

The contributions of the present work are summarized in the following theorems, which indicate that the aforedescribed surprising phenomena can in fact be observed for all binary-input channels.

\begin{theorem}\label{thm:theorem1}
For any binary-input channel $p_{Y|X,S}$, state distribution $p_S$, and side channel $p_{\tilde{S}|S}$,
\begin{align*}
C(p_{Y|X,S},p_S,p_{\tilde{S}|S})=\underline{C}(p_{Y|X,S},p_S)
\end{align*}
if $I(S;\tilde{S})\leq\frac{\rho^2}{2e^2}$, where $\rho\triangleq\min_{s\in\mathcal{S}}p_S(s)$.
\end{theorem}

\begin{theorem}\label{thm:theorem2}
For any binary-input channel $p_{Y|X,S}$, state distribution $p_S$, and side channel $p_{\tilde{S}|S}$,
\begin{align*}
C'(p_{Y|X,S},p_S,p_{\tilde{S}|S})=\overline{C}(p_{Y|X,S},p_S)
\end{align*}
if  $H(S|\tilde{S})\leq\frac{2\rho\log 2}{(|\mathcal{S}|-1)(e-1)}$, where $\rho\triangleq\min_{s\in\mathcal{S}}p_S(s)$.
\end{theorem}

On the surface these two results may look rather similar. One might even suspect the existence of a certain duality between them. However, it will be seen that the underlying reasons are actually quite different. The proof of Theorem \ref{thm:theorem1} hinges upon, among other things, a perturbation analysis. In contrast, Theorem \ref{thm:theorem2} is essentially a manifestation of an induced Markov structure.

The conditions in Theorem \ref{thm:theorem1} and Theorem \ref{thm:theorem2} are stated in terms of bounds on $I(S;\tilde{S})$ and $H(S|\tilde{S})$; as a consequence, they depend inevitably on $p_S$. As shown by Theorem \ref{thm:theorem1variant} in Section \ref{sec:proof1} and Theorem \ref{thm:theorem2variant} in Section \ref{sec:proof2}, it is in fact possible to establish these two results under more general conditions on $p_{\tilde{S}|S}$ that are universal for all binary-input channels and state distributions.

The rest of this paper is organized as follows. We present the proofs of Theorems \ref{thm:theorem1} and \ref{thm:theorem2} in Sections \ref{sec:proof1} and \ref{sec:proof2}, respectively. The validity of these two results under various modified conditions is discussed in Section \ref{sec:discussion}. Section \ref{sec:conclusion} contains some concluding remarks. Throughout this paper, all logarithms are base-$e$.


\section{Proof of Theorem \ref{thm:theorem1}}\label{sec:proof1}


First consider the special case where $p_{\tilde{S}|S}$ is a generalized erasure channel (with erasure probability $\epsilon\in[0,1]$)
defined as
\begin{align*}
p_{\tilde{S}^{(\epsilon)}_{GE}|S}(\tilde{s}|s)=\left\{
                               \begin{array}{ll}
                                 1-\epsilon, & \tilde{s}=s, \\
                                 \epsilon, & \tilde{s}=\ast, \\
                                 0, & \mbox{otherwise},
                               \end{array}
                             \right.\quad s\in\mathcal{S}, \tilde{s}\in\mathcal{S}\cup\{*\}.
\end{align*}

\begin{lemma}\label{lem:core}
Given any binary-input channel $p_{Y|X,S}$ and state distribution $p_S$,
\begin{align*}
C(p_{Y|X,S},p_S,p_{\tilde{S}^{(\epsilon)}_{GE}|S})=\underline{C}(p_{Y|X,S},p_S)
\end{align*}
for $\epsilon\in[1-e^{-1},1]$.
\end{lemma}
\emph{Remark:} Lemma \ref{lem:core} provides a universal upper bound\footnote{Numerical simulations suggest that this universal upper bound is not tight. Determining the exact universal erasure probability threshold remains an open problem.} on the erasure probability threshold above which the encoder side information is useless. The actual threshold, however, depends on $p_{Y|X,S}$ and $p_S$ (see Section \ref{subsec:extensionthm1} for a detailed analysis).
\begin{IEEEproof}
As indicated by (\ref{eq:noisystate}), the capacity of the channel model in Fig. \ref{fig:model} (i.e., $C(p_{Y|X,S},p_S,p_{\tilde{S}|S})$) is equal to that of channel $p_{Y,S|U}$, where
\begin{align*}
&p_{Y,S|U}(y,s|u)=\sum\limits_{\tilde{s}\in\tilde{\mathcal{S}}}p_S(s)p_{\tilde{S}|S}(\tilde{s}|s)p_{Y|X,S}(y|\psi(u,\tilde{s}),s),\\
&\hspace{2.2in} u\in\mathcal{U},y\in\mathcal{Y},s\in\mathcal{S}.
\end{align*}
According to \cite[Th. 4.5.1]{Gallager68}, $p_U$ is a capacity-achieving input distribution of channel $p_{Y,S|U}$  (i.e., $p_{U}$ is a maximizer of the optimization problem in (\ref{eq:noisystate})) if and only if there exists some number $C$ such that
\begin{align*}
&D(p_{Y,S|U}(\cdot,\cdot|u)\|p_{Y,S})=C,\quad u\in\mathcal{U}\mbox{ with }p_U(u)>0, \\
&D(p_{Y,S|U}(\cdot,\cdot|u)\|p_{Y,S})\leq C,\quad u\in\mathcal{U}\mbox{ with }p_U(u)=0;
\end{align*}
furthermore, the number $C$ is equal to $C(p_{Y|X,S},p_S,p_{\tilde{S}|S})$.
In view of (\ref{eq:orderpsi}), we have
\begin{align*}
p_{Y,S|U}(y,s|u)=p_{Y,S|X}(y,s|u),\quad u\in\mathcal{X},y\in\mathcal{Y},s\in\mathcal{S}.
\end{align*}
Let $p_{\hat{X}}$ be a capacity-achieving input distribution of channel $p_{Y,S|X}$ (i.e, $p_{\hat{X}}$ is a maximizer of the optimization problem in (\ref{eq:nostate})). Define
\begin{align}
p_{\hat{U}}(u)=\left\{
                 \begin{array}{ll}
                   p_{\hat{X}}(u), & u\in\mathcal{X}, \\
                   0, & \mbox{otherwise}.
                 \end{array}
               \right.\label{eq:uconstruction}
\end{align}
It is clear that $C(p_{Y|X,S},p_S,p_{\tilde{S}|S})=\underline{C}(p_{Y|X,S},p_S)$ if and only if $p_{\hat{U}}$ is a capacity-achieving input distribution of channel $p_{Y,S|U}$.

Now consider the special case where $p_{\tilde{S}|S}$ is a generalized erasure channel with erasure probability $\epsilon$, and define
\begin{align}
D_{GE}(p_U,\epsilon,u)=D(p_{Y,S|U}(\cdot,\cdot|u)\|p_{Y,S})\label{eq:defD}
\end{align}
to stress the dependence of $D(p_{Y,S|U}(\cdot,\cdot|u)\|p_{Y,S})$ on $p_U$, $\epsilon$, and $u$. It can be verified that
\begin{align}
&p_{Y,S|U}(y,s|u)\nonumber\\
&=\sum\limits_{\tilde{s}\in\mathcal{S}\cup\{\ast\}}p_S(s)p_{\tilde{S}^{(\epsilon)}|S}(\tilde{s}|s)p_{Y|X,S}(y|\psi(u,\tilde{s}),s)\nonumber\\
&=p_S(s)\epsilon p_{Y|X,S}(y|\psi(u,\ast),s)\nonumber\\
&\quad+p_S(s)(1-\epsilon)p_{Y|X,S}(y|\psi(u,s),s)\nonumber\\
&=p_S(s)(p_{Y|X,S}(y|\psi(u,s),s)+\epsilon\delta(u,y,s)),\label{eq:subH2}
\end{align}
where
\begin{align}
&\delta(u,y,s)=p_{Y|X,S}(y|\psi(u,\ast),s)-p_{Y|X,S}(y|\psi(u,s),s),\nonumber\\
&\hspace{1.8in}u\in\mathcal{U},y\in\mathcal{Y},s\in\mathcal{S}.\label{eq:defdelta}
\end{align}
Since $|\mathcal{X}|=2$, there is no loss of generality  in assuming that \cite[Th. 2]{SF04}
\begin{align}
p_{\hat{X}}(x)>e^{-1},\quad x\in\mathcal{X}.\label{eq:posinput}
\end{align}
To the end of proving Lemma \ref{lem:core}, it suffices to show that, for $\epsilon\in[1-e^{-1},1]$,
\begin{align*}
&D_{GE}(p_{\hat{U}},\epsilon,u)=\underline{C}(p_{Y|X,S},p_S),\quad u\in\mathcal{X}, \\
&D_{GE}(p_{\hat{U}},\epsilon,u)\leq \underline{C}(p_{Y|X,S},p_S),\quad \mbox{otherwise}.
\end{align*}

Clearly, $p_{\hat{U}}$ is a capacity-achieving input distribution of channel $p_{Y,S|U}$ when $\epsilon=1$. Therefore, we have\footnote{The inequality in (\ref{eq:ineqtoeq}) is in fact an equality.}
\begin{align}
&D_{GE}(p_{\hat{U}},1,u)=\underline{C}(p_{Y|X,S},p_S),\quad u\in\mathcal{X},\label{eq:twoactiveinputs}\\
&D_{GE}(p_{\hat{U}},1,u)\leq\underline{C}(p_{Y|X,S},p_S),\quad \mbox{otherwise}.\label{eq:ineqtoeq}
\end{align}
Note that
\begin{align}
&D_{GE}(p_{\hat{U}},\epsilon,u)\nonumber\\
&=\sum\limits_{y\in\mathcal{Y},s\in\mathcal{S}} p_{Y,S|U}(y,s|u)\log\frac{p_{Y,S|U}(y,s|u)}{\sum_{u'\in\mathcal{U}}p_{\hat{U}}(u')p_{Y,S|U}(y,s|u')}\nonumber\\
&=\sum\limits_{y\in\mathcal{Y},s\in\mathcal{S}} p_S(s)(p_{Y|X,S}(y|\psi(u,s),s)+\epsilon\delta(u,y,s))\nonumber\\
&\quad\times\log\frac{p_{Y|X,S}(y|\psi(u,s),s)+\epsilon\delta(u,y,s)}{\sum_{u'\in\mathcal{U}}p_{\hat{U}}(u')(p_{Y|X,S}(y|\psi(u',s),s)+\epsilon\delta(u',y,s))}\label{eq:subp}\\
&=\sum\limits_{y\in\mathcal{Y},s\in\mathcal{S}} p_S(s)(p_{Y|X,S}(y|\psi(u,s),s)+\epsilon\delta(u,y,s))\nonumber\\
&\hspace{0.6in}\times\log\frac{p_{Y|X,S}(y|\psi(u,s),s)+\epsilon\delta(u,y,s)}{\sum_{x\in\mathcal{X}}p_{\hat{X}}(x)p_{Y|X,S}(y|x,s)},\nonumber\\
&\hspace{2.0in}\epsilon\in[0,1], u\in\mathcal{U},\label{eq:combine3}
\end{align}
where (\ref{eq:subp}) is due to (\ref{eq:subH2}), and (\ref{eq:combine3}) is due to (\ref{eq:orderpsi}) and (\ref{eq:uconstruction}). Moreover,
\begin{align}
&\frac{\partial}{\partial\epsilon}D_{GE}(p_{\hat{U}},\epsilon,u)\nonumber\\
&=\sum\limits_{y\in\mathcal{Y},s\in\mathcal{S}} p_S(s)\delta(u,y,s) \nonumber\\
&\hspace{0.5in}\times\Bigg(\log\frac{p_{Y|X,S}(y|\psi(u,s),s)+\epsilon\delta(u,y,s)}{\sum_{x\in\mathcal{X}}p_{\hat{X}}(x)p_{Y|X,S}(y|x,s)}+1\Bigg)\nonumber\\
&=\sum\limits_{y\in\mathcal{Y},s\in\mathcal{S}} p_S(s)\delta(u,y,s)\nonumber\\
&\hspace{0.5in}\times\log\frac{p_{Y|X,S}(y|\psi(u,s),s)+\epsilon\delta(u,y,s)}{\sum_{x\in\mathcal{X}}p_{\hat{X}}(x)p_{Y|X,S}(y|x,s)}\nonumber\\
&\quad+\sum\limits_{s\in\mathcal{S}}p_S(s)\sum\limits_{y\in\mathcal{Y}}\delta(u,y,s)\nonumber\\
&=\sum\limits_{y\in\mathcal{Y},s\in\mathcal{S}} p_S(s)\delta(u,y,s)\nonumber\\
&\hspace{0.5in}\times\log\frac{p_{Y|X,S}(y|\psi(u,s),s)+\epsilon\delta(u,y,s)}{\sum_{x\in\mathcal{X}}p_{\hat{X}}(x)p_{Y|X,S}(y|x,s)},\nonumber\\
&\hspace{2in} \epsilon\in[0,1], u\in\mathcal{U}.\label{eq:newsub}
\end{align}
Define
\begin{align}
\mathcal{G}_{\delta}=\{u\in\mathcal{U}: \delta(u,y,s)=0\mbox{ for all }y\in\mathcal{Y}\mbox{ and }s\in\mathcal{S}\}.\label{eq:defutilde}
\end{align}
In light of (\ref{eq:combine3}),
\begin{align}
D_{GE}(p_{\hat{U}},\epsilon,u)=D_{GE}(p_{\hat{U}},1,u),\quad \epsilon\in[0,1], u\in\mathcal{G}_{\delta}.\label{eq:invariant}
\end{align}
For any $u\in\mathcal{U}\backslash\mathcal{G}_{\delta}$, there must exist some $y\in\mathcal{Y}$ and $s\in\mathcal{S}$ such that $\delta(u,y,s)\neq 0$; furthermore, since $|\mathcal{X}|=2$, we have
\begin{align}
\delta(u,y,s)>0&\Longrightarrow   p_{Y|X,S}(y|\psi(u,s),s)+\epsilon\delta(u,y,s)\nonumber\\
&\qquad=b(y,s)+\epsilon(a(y,s)-b(y,s)),\label{eq:pos}\\
\delta(u,y,s)<0&\Longrightarrow   p_{Y|X,S}(y|\psi(u,s),s)+\epsilon\delta(u,y,s)\nonumber\\
&\qquad=a(y,s)+\epsilon(b(y,s)-a(y,s)),\label{eq:neg}
\end{align}
where
\begin{align*}
&a(y,s)=\max\limits_{x\in\mathcal{X}}p_{Y|X,S}(y|x,s),\\
&b(y,s)=\min\limits_{x\in\mathcal{X}}p_{Y|X,S}(y|x,s).
\end{align*}
Continuing from (\ref{eq:newsub}),
\begin{align}
&\frac{\partial}{\partial\epsilon}D_{GE}(p_{\hat{U}},\epsilon,u)\nonumber\\
&=\sum\limits_{y\in\mathcal{Y},s\in\mathcal{S}} p_S(s)\delta(u,y,s)\nonumber\\
&\hspace{0.5in}\times\log\frac{p_{Y|X,S}(y|\psi(u,s),s)+\epsilon\delta(u,y,s)}{\sum_{x\in\mathcal{X}}p_{\hat{X}}(x)p_{Y|X,S}(y|x,s)}\nonumber\\
&\geq\sum\limits_{s\in\mathcal{S}}p_S(s)\sum\limits_{y\in\mathcal{Y}:\delta(u,y,s)>0}\delta(u,y,s)\nonumber\\
&\hspace{0.5in}\times\log\frac{p_{Y|X,S}(y|\psi(u,s),s)+\epsilon\delta(u,y,s)}{(1-e^{-1})a(y,s)+e^{-1}b(y,s)}\nonumber\\
&\quad+\sum\limits_{s\in\mathcal{S}}p_S(s)\sum\limits_{y\in\mathcal{Y}:\delta(u,y,s)<0}\delta(u,y,s)\nonumber\\
&\hspace{0.5in}\times\log\frac{p_{Y|X,S}(y|\psi(u,s),s)+\epsilon\delta(u,y,s)}{e^{-1}a(y,s)+(1-e^{-1})b(y,s)}\label{eq:invokeposinput}\\
&=\sum\limits_{s\in\mathcal{S}}p_S(s)\sum\limits_{y\in\mathcal{Y}:\delta(u,y,s)>0}\delta(u,y,s)\nonumber\\
&\hspace{0.5in}\times\log\frac{b(y,s)+\epsilon(a(y,s)-b(y,s))}{(1-e^{-1})a(y,s)+e^{-1}b(y,s)}\nonumber\\
&\quad+\sum\limits_{s\in\mathcal{S}}p_S(s)\sum\limits_{y\in\mathcal{Y}:\delta(u,y,s)<0}\delta(u,y,s)\nonumber\\
&\hspace{0.5in}\times\log\frac{a(y,s)+\epsilon(b(y,s)-a(y,s))}{e^{-1}a(y,s)+(1-e^{-1})b(y,s)}\label{eq:twoimply}\\
&\geq\sum\limits_{s\in\mathcal{S}}p_S(s)\sum\limits_{y\in\mathcal{Y}:\delta(u,y,s)>0}\delta(u,y,s)\nonumber\\
&\hspace{0.5in}\times\log\frac{(1-e^{-1}) a(y,s)+e^{-1}b(y,s)}{(1-e^{-1})a(y,s)+e^{-1}b(y,s)}\nonumber\\
&\quad+\sum\limits_{s\in\mathcal{S}}p_S(s)\sum\limits_{y\in\mathcal{Y}:\delta(u,y,s)<0}\delta(u,y,s)\nonumber\\
&\hspace{0.5in}\times\log\frac{e^{-1}a(y,s)+(1-e^{-1}) b(y,s)}{e^{-1}a(y,s)+(1-e^{-1})b(y,s)}\nonumber\\
&=0,\quad \epsilon\in[1-e^{-1},1], u\in\mathcal{U},\label{eq:comb4}
\end{align}
where (\ref{eq:invokeposinput}) is due to (\ref{eq:posinput}), and (\ref{eq:twoimply}) is due to (\ref{eq:pos}) and (\ref{eq:neg}).
Combining (\ref{eq:twoactiveinputs}), (\ref{eq:ineqtoeq}), (\ref{eq:invariant}), (\ref{eq:comb4}), and the fact $\mathcal{X}\subseteq\mathcal{G}_{\delta}$  yields the desired result.
\end{IEEEproof}

Recall \cite[p. 112]{EGK11} that $p_{\tilde{S}_1|S}$ (with input alphabet $\mathcal{S}$ and output alphabet $\tilde{\mathcal{S}}_1$) is said to be a stochastically degraded version of $p_{\tilde{S}_2|S}$ (with input alphabet $\mathcal{S}$ and output alphabet $\tilde{\mathcal{S}}_2$) if there exists $p_{\tilde{S}_1|\tilde{S}_2}$ satisfying
\begin{align}
&p_{\tilde{S}_1|S}(\tilde{s}_1|s)=\sum\limits_{\tilde{s}_2\in\tilde{\mathcal{S}}_2}p_{\tilde{S}_2|S}(\tilde{s}_2|s)p_{\tilde{S}_1|\tilde{S}_2}(\tilde{s}_1|\tilde{s}_2),\nonumber\\
&\hspace{2in} s\in\mathcal{S},\tilde{s}_1\in\tilde{\mathcal{S}}_1.\label{eq:equivalent}
\end{align}
We can write (\ref{eq:equivalent}) equivalently as
\begin{align*}
p_{\tilde{S}_1|S}=p_{\tilde{S}_2|S}p_{\tilde{S}_1|\tilde{S}_2}
\end{align*}
by viewing $p_{\tilde{S}_1|S}$, $p_{\tilde{S}_2|S}$, and $p_{\tilde{S}_1|\tilde{S}_2}$ as probability transition matrices.

The following result is obvious and its proof is omitted. 

\begin{lemma}\label{lem:degraded}
If $p_{\tilde{S}_1|S}$ is a stochastically degraded version of $p_{\tilde{S}_2|S}$, then
\begin{align*}
C(p_{Y|X,S},p_S,p_{\tilde{S}_1|S})\leq C(p_{Y|X,S},p_S,p_{\tilde{S}_2|S}).
\end{align*}
\end{lemma}

Next we extend Lemma \ref{lem:core} to the general case by characterizing the condition under which $p_{\tilde{S}|S}$ is a stochastically degraded  version of $p_{\tilde{S}^{(\epsilon)}_{GE}|S}$.

\begin{lemma}\label{lem:erasure}
$p_{\tilde{S}|S}$ is a stochastically degraded version of  $p_{\tilde{S}^{(\epsilon)}_{GE}|S}$ if and only if
\begin{align}
\sum\limits_{\tilde{s}\in\tilde{S}}\min\limits_{s\in\mathcal{S}}p_{\tilde{S}|S}(\tilde{s}|s)\geq\epsilon.\label{eq:ns}
\end{align}
\end{lemma}
\begin{IEEEproof}
The problem boils down to finding a necessary and sufficient condition for the existence of $p_{\tilde{S}|\tilde{S}^{(\epsilon)}_{GE}}$ such that
\begin{align}
&p_{\tilde{S}|S}(\tilde{s}|s)=\sum\limits_{\tilde{s}'\in\mathcal{S}\cup\{\ast\}}p_{\tilde{S}^{(\epsilon)}_{GE}|S}(\tilde{s}'|s)p_{\tilde{S}|\tilde{S}^{(\epsilon)}_{GE}}(\tilde{s}|\tilde{s}'),\nonumber\\
&\hspace{2in} s\in\mathcal{S}, \tilde{s}\in\tilde{\mathcal{S}}.\label{eq:cond1}
\end{align}
It suffices to consider the case $\epsilon\in[0,1)$ since Lemma \ref{lem:erasure} is trivially true when $\epsilon=1$. Note that
\begin{align}
&\sum\limits_{\tilde{s}'\in\mathcal{S}\cup\{\ast\}}p_{\tilde{S}^{(\epsilon)}_{GE}|S}(\tilde{s}'|s)p_{\tilde{S}|\tilde{S}^{(\epsilon)}_{GE}}(\tilde{s}|\tilde{s}')\nonumber\\
&=(1-\epsilon)p_{\tilde{S}|\tilde{S}^{(\epsilon)}_{GE}}(\tilde{s}|s)+\epsilon p_{\tilde{S}|\tilde{S}^{(\epsilon)}_{GE}}(\tilde{s}|\ast),\quad s\in\mathcal{S}, \tilde{s}\in\tilde{\mathcal{S}}.\label{eq:cond2}
\end{align}
Combining (\ref{eq:cond1}) and (\ref{eq:cond2}) gives
\begin{align}
p_{\tilde{S}|\tilde{S}^{(\epsilon)}_{GE}}(\tilde{s}|s)=\frac{p_{\tilde{S}|S}(\tilde{s}|s)-\epsilon p_{\tilde{S}|\tilde{S}^{(\epsilon)}_{GE}}(\tilde{s}|\ast)}{1-\epsilon}, \quad s\in\mathcal{S}, \tilde{s}\in\tilde{\mathcal{S}}.\label{eq:inview}
\end{align}
In light of (\ref{eq:inview}),
\begin{align}
&\sum\limits_{\tilde{s}\in\tilde{\mathcal{S}}} p_{\tilde{S}|\tilde{S}^{(\epsilon)}_{GE}}(\tilde{s}|s)=1,\quad s\in\mathcal{S},\nonumber\\
&\Longleftrightarrow\sum\limits_{\tilde{s}\in\tilde{\mathcal{S}}}p_{\tilde{S}|\tilde{S}^{(\epsilon)}_{GE}}(\tilde{s}|\ast)=1,\nonumber\\
&p_{\tilde{S}|\tilde{S}^{(\epsilon)}_{GE}}(\tilde{s}|s)\geq 0,\quad s\in\mathcal{S}, \tilde{s}\in\tilde{\mathcal{S}},\nonumber\\
&\Longleftrightarrow \min\limits_{s\in\mathcal{S}}p_{\tilde{S}|S}(\tilde{s}|s)\geq\epsilon p_{\tilde{S}|\tilde{S}^{(\epsilon)}_{GE}}(\tilde{s}|\ast),\quad \tilde{s}\in\tilde{\mathcal{S}}.\label{eq:ns2}
\end{align}
It can be readily seen that the existence of conditional distribution $p_{\tilde{S}|\tilde{S}^{(\epsilon)}_{GE}}$ satisfying (\ref{eq:cond1}) is equivalent to the existence of probability vector $(p_{\tilde{S}|\tilde{S}^{(\epsilon)}_{GE}}(\tilde{s}|\ast))_{\tilde{s}\in\tilde{\mathcal{S}}}$ satisfying (\ref{eq:ns2}). Clearly,  (\ref{eq:ns}) is a necessary and sufficient condition for the existence of such $(p_{\tilde{S}|\tilde{S}^{(\epsilon)}_{GE}}(\tilde{s}|\ast))_{\tilde{s}\in\tilde{\mathcal{S}}}$.
\end{IEEEproof}

\begin{theorem}\label{thm:theorem1variant}
For any binary-input channel $p_{Y|X,S}$, state distribution $p_S$, and side channel $p_{\tilde{S}|S}$,
\begin{align*}
C(p_{Y|X,S},p_S,p_{\tilde{S}|S})=\underline{C}(p_{Y|X,S},p_S)
\end{align*}
if
\begin{align}
\sum\limits_{\tilde{s}\in\tilde{S}}\min\limits_{s\in\mathcal{S}}p_{\tilde{S}|S}(\tilde{s}|s)\geq 1-e^{-1}.\label{eq:translate}
\end{align}
\end{theorem}
\begin{IEEEproof}
In view of Lemmas \ref{lem:core}, \ref{lem:degraded}, and \ref{lem:erasure}, we have
\begin{align}
C(p_{Y|X,S},p_S,p_{\tilde{S}|S})\leq\underline{C}(p_{Y|X,S},p_S)\label{eq:comb2}
\end{align}
if (\ref{eq:translate}) is satisfied.
Combining (\ref{eq:comb1}) and (\ref{eq:comb2}) completes the proof of Theorem \ref{thm:theorem1variant}.
\end{IEEEproof}

Now we proceed to prove Theorem \ref{thm:theorem1} by translating (\ref{eq:translate}) (which is a condition on $p_{\tilde{S}|S}$ that is universal for all binary input channels and state distributions) to an upper bound on $I(S;\tilde{S})$. This upper bound, however, depends inevitably on the state distribution.

For any $p_{\tilde{S}|S}$ violating (\ref{eq:translate}) (i.e, $\sum_{\tilde{s}\in\tilde{S}}\min_{s\in\mathcal{S}}p_{\tilde{S}|S}(\tilde{s}|s)<1-e^{-1}$), we have
\begin{align}
I(S;\tilde{S})&\geq\frac{1}{2}\Bigg(\sum\limits_{s\in\mathcal{S},\tilde{s}\in\tilde{\mathcal{S}}}p_S(s)\left|p_{\tilde{S}}(\tilde{s})-p_{\tilde{S}|S}(\tilde{s}|s)\right|\Bigg)^2\label{eq:pinsker}\\
&\geq\frac{1}{2}\Bigg(\sum\limits_{\tilde{s}\in\tilde{\mathcal{S}}}p_S(s(\tilde{s}))\left|p_{\tilde{S}}(\tilde{s})-p_{\tilde{S}|S}(\tilde{s}|s(\tilde{s}))\right|\Bigg)^2\nonumber\\
&\geq\frac{1}{2}\Bigg(\rho\sum\limits_{\tilde{s}\in\tilde{\mathcal{S}}}\left|p_{\tilde{S}}(\tilde{s})-p_{\tilde{S}|S}(\tilde{s}|s(\tilde{s}))\right|\Bigg)^2\nonumber\\
&\geq\frac{1}{2}\Bigg(\rho\left|\sum\limits_{\tilde{s}\in\tilde{\mathcal{S}}}p_{\tilde{S}}(\tilde{s})-\sum\limits_{\tilde{s}\in\tilde{\mathcal{S}}}p_{\tilde{S}|S}(\tilde{s}|s(\tilde{s}))\right|\Bigg)^2\nonumber\\
&>\frac{\rho^2}{2e^2},\nonumber
\end{align}
where (\ref{eq:pinsker}) is due to Pinsker's inequality \cite[p. 44]{CK01B}, and $s(\tilde{s})$ is a minimizer of $\min_{s\in\mathcal{S}}p_{\tilde{S}|S}(\tilde{s}|s)$, $\tilde{s}\in\tilde{\mathcal{S}}$. As a consequence, (\ref{eq:translate}) must hold if $I(S;\tilde{S})\leq\frac{\rho^2}{2e^2}$. This completes the proof of Theorem \ref{thm:theorem1}.


\section{Proof of Theorem \ref{thm:theorem2}}\label{sec:proof2}

First consider the special case where $p_{\tilde{S}|S}$ is a generalized symmetric channel (with crossover probability $q\in[0,\frac{1}{|\mathcal{S}|}]$)
defined as
\begin{align*}
p_{\tilde{S}^{(q)}_{GS}|S}(\tilde{s}|s)=\left\{
                               \begin{array}{ll}
                                 1-(|\mathcal{S}|-1)q, & \tilde{s}=s, \\
                                 q, & \mbox{otherwise},
                               \end{array}
                             \right.\hspace{0.03in} s\in\mathcal{S}, \tilde{s}\in\mathcal{S}.
\end{align*}

\begin{lemma}\label{lem:nssym}
$C'(p_{Y|X,S},p_S,p_{\tilde{S}^{(q)}_{GS}|S})=\overline{C}(p_{Y|X,S},p_S)$ if and only if
\begin{align}
\min\limits_{x\in\mathcal{X}_+,s\in\mathcal{S}}\frac{p_{\hat{X}|S}(x|s)}{\sum_{s'\in\mathcal{S}}p_{\hat{X}|S}(x|s')}\geq q\label{eq:nssym}
\end{align}
for some $p_{\hat{X}|S}\in\mathcal{P}$, where $\mathcal{P}$ denotes the set of maximizers of the optimization problem in (\ref{eq:perfectstate}), and $\mathcal{X}_+=\{x\in\mathcal{X}:\sum_{s\in\mathcal{S}}p_{\hat{X}|S}(x|s)>0\}$.
\end{lemma}
\begin{IEEEproof}
Clearly, $C'(p_{Y|X,S},p_S,p_{\tilde{S}^{(q)}_{GS}|S})=\overline{C}(p_{Y|X,S},p_S)$ if and only if there exists $p_{\hat{X}|S}\in\mathcal{P}$ that is a stochastically degraded version of $p_{\tilde{S}^{(q)}_{GS}|S}$. When $q=\frac{1}{|\mathcal{S}|}$, (\ref{eq:nssym}) is equivalent to the desired condition that $\hat{X}$ needs to be independent of $S$.
When $q\in[0,\frac{1}{|\mathcal{S}|})$, $p_{\tilde{S}^{(q)}_{GS}|S}$ is invertible and
\begin{align}
p^{-1}_{\tilde{S}^{(q)}_{GS}|S}&=\left(
                                                \begin{array}{cccc}
                                                  \frac{q-1}{|\mathcal{S}|q-1} &\frac{q}{|\mathcal{S}|q-1} & \cdots & \frac{q}{|\mathcal{S}|q-1} \\
                                                   \frac{q}{|\mathcal{S}|q-1} & \ddots & \ddots & \vdots\\
                                                  \vdots & \ddots & \ddots & \frac{q}{|\mathcal{S}|q-1} \\
                                                  \frac{q}{|\mathcal{S}|q-1} & \cdots & \frac{q}{|\mathcal{S}|q-1} & \frac{q-1}{|\mathcal{S}|q-1} \\
                                                \end{array}
                                              \right).\label{eq:inverse}
\end{align}
The problem boils down to finding a necessary and sufficient condition under which $p^{-1}_{\tilde{S}^{(q)}_{GS}|S}p_{\hat{X}|S}$ is a valid probability transition matrix (i.e., all entries are non-negative and the sum of each row vector is equal to 1). Note that
\begin{align}
p^{-1}_{\tilde{S}^{(q)}_{GS}|S}p_{\hat{X}|S}\left(
                                     \begin{array}{c}
                                       1 \\
                                       \vdots \\
                                       1 \\
                                     \end{array}
                                   \right)&=p^{-1}_{\tilde{S}^{(q)}_{GS}|S}\left(
                                     \begin{array}{c}
                                       1 \\
1\\
                                       \vdots \\
                                       1 \\
                                     \end{array}
                                   \right)\nonumber\\
&=p^{-1}_{\tilde{S}^{(q)}_{GS}|S}p_{\tilde{S}^{(q)}_{GS}|S}\left(
                                     \begin{array}{c}
                                       1 \\
                                       1\\
                                       \vdots \\
                                       1 \\
                                     \end{array}
                                   \right)\nonumber\\
&=\left(
                                     \begin{array}{c}
                                       1 \\
1\\
                                       \vdots \\
                                       1 \\
                                     \end{array}
                                   \right).\label{eq:identity}
\end{align}
Moreover, all entries of $p^{-1}_{\tilde{S}^{(q)}_{GS}|S}p_{\hat{X}|S}$ are non-negative if and only if
\begin{align*}
\frac{-p_{\hat{X}|S}(x|s)+q\sum_{s'\in\mathcal{S}}p_{\hat{X}|S}(x|s')}{|\mathcal{S}|q-1}\geq 0,\quad x\in\mathcal{S}, s\in\mathcal{S},
\end{align*}
which is equivalent to (\ref{eq:nssym}).
\end{IEEEproof}

The following result is obvious and its proof is omitted.
\begin{lemma}\label{lem:degraded2}
If $p_{\tilde{S}_1|S}$ is a stochastically degraded version of $p_{\tilde{S}_2|S}$, then
\begin{align*}
C'(p_{Y|X,S},p_S,p_{\tilde{S}_1|S})\leq C'(p_{Y|X,S},p_S,p_{\tilde{S}_2|S}).
\end{align*}
\end{lemma}

\begin{lemma}\label{lem:Mmatrix}
$p_{\tilde{S}^{(q)}_{GS}|S}$ is a stochastically degraded version of $p_{\tilde{S}|S}$ if
\begin{align}
\max\limits_{s\in\mathcal{S},\hat{s}\in\mathcal{S}_+:s\neq\hat{s}}\frac{p_{\hat{S}|S}(\hat{s}|s)}{\sum_{s'\in\mathcal{S}}p_{\hat{S}|S}(\hat{s}|s')}\leq q,\label{eq:suff}
\end{align}
where $\hat{S}$ is the maximum likelihood estimate of $S$ based on $\tilde{S}$, and $\mathcal{S}_+=\{\hat{s}\in\mathcal{S}:\sum_{s\in\mathcal{S}}p_{\hat{S}|S}(\hat{s}|s)>0\}$.
\end{lemma}
\begin{IEEEproof}
The case $q=\frac{1}{|\mathcal{S}|}$ is trivial. When $q\in[0,\frac{1}{|\mathcal{S}|})$, $p_{\tilde{S}^{(q)}_{GS}|S}$ is invertible and $p^{-1}_{\tilde{S}^{(q)}_{GS}|S}$ is given by (\ref{eq:inverse}). It can be shown (see the derivation of (\ref{eq:identity})) that the sum of each row of $p^{-1}_{\tilde{S}^{(q)}_{GS}|S}p_{\hat{S}|S}$ is equal to 1; moreover, the off-diagonal entries of $p^{-1}_{\tilde{S}^{(q)}_{GS}|S}p_{\hat{S}|S}$ are non-positive if and only if
\begin{align*}
&\frac{-p_{\hat{S}|S}(\hat{s}|s)+q\sum_{s'\in\mathcal{S}}p_{\hat{S}|S}(\hat{s}|s')}{|\mathcal{S}|q-1}\leq 0,\\
&\hspace{1.5in} s\in\mathcal{S},\hat{s}\in\mathcal{S}_+:s\neq\hat{s},
\end{align*}
which is equivalent to (\ref{eq:suff}). Therefore,  (\ref{eq:suff}) ensures that $p^{-1}_{\tilde{S}^{(q)}_{GS}|S}p_{\hat{S}|S}$ is a non-singular $M$-matrix, which in turn ensures that $p^{-1}_{\hat{S}|S}p_{\tilde{S}^{(q)}_{GS}|S}$ exists and is a non-negative matrix \cite{Plemmons77}. Hence, if (\ref{eq:suff}) is satisfied, then $p^{-1}_{\hat{S}|S}p_{\tilde{S}^{(q)}_{GS}|S}$  is a valid probability transition matrix (the requirement that the entries in each row of $p^{-1}_{\hat{S}|S}p_{\tilde{S}^{(q)}_{GS}|S}$ add up to 1 is automatically satisfied), which implies that
$p_{\tilde{S}^{(q)}_{GS}|S}$ is a stochastically degraded version of $p_{\hat{S}|S}$ (and consequently a stochastically degraded version of $p_{\tilde{S}|S}$).
\end{IEEEproof}

\begin{theorem}\label{thm:theorem2variant}
For any binary-input channel $p_{Y|X,S}$, state distribution $p_S$, and side channel $p_{\tilde{S}|S}$,
\begin{align*}
C'(p_{Y|X,S},p_S,p_{\tilde{S}|S})=\overline{C}(p_{Y|X,S},p_S)
\end{align*}
if
\begin{align}
\max\limits_{s\in\mathcal{S},\hat{s}\in\mathcal{S}_+:s\neq\hat{s}}\frac{p_{\hat{S}|S}(\hat{s}|s)}{\sum_{s'\in\mathcal{S}}p_{\hat{S}|S}(\hat{s}|s')}\leq\frac{1}{(|\mathcal{S}|-1)e-|\mathcal{S}|+2},\label{eq:sidechannelcondition}
\end{align}
where $\hat{S}$ is the maximum likelihood estimate of $S$ based on $\tilde{S}$.
\end{theorem}
\begin{IEEEproof}
Since $|\mathcal{X}|=2$, it follows from \cite[Th. 2]{SF04} that there exists $p_{\hat{X}|S}\in\mathcal{P}$ satisfying
\begin{align*}
p_{\hat{X}|S}(x|s)>e^{-1},\quad x\in\mathcal{X}, s\in\mathcal{S}.
\end{align*}
For such $p_{\hat{X}|S}$,
\begin{align*}
\min\limits_{x\in\mathcal{X}_+,s\in\mathcal{S}}\frac{p_{\hat{X}|S}(x|s)}{\sum_{s'\in\mathcal{S}}p_{\hat{X}|S}(x|s')}&\geq\frac{e^{-1}}{e^{-1}+(|\mathcal{S}|-1)(1-e^{-1})}\\
&=\frac{1}{(|\mathcal{S}|-1)e-|\mathcal{S}|+2}.
\end{align*}
In view of of Lemmas \ref{lem:nssym}, \ref{lem:degraded2}, and \ref{lem:Mmatrix}, we have
\begin{align}
C'(p_{Y|X,S},p_S,p_{\tilde{S}|S})\geq\overline{C}(p_{Y|X,S},p_S)\label{eq:comb3}
\end{align}
if (\ref{eq:sidechannelcondition}) is satisfied. Combining (\ref{eq:comb1}) and (\ref{eq:comb3}) completes the proof of Theorem \ref{thm:theorem2variant}.
\end{IEEEproof}


Now we are in a position to prove Theorem \ref{thm:theorem2}.  Let $\hat{S}$ and $\hat{S}'$ denote respectively the maximum likelihood estimate and the maximum \textit{a posteriori} estimate of $S$ based on $\tilde{S}$. 
According to \cite[Th. 11]{HV10},
\begin{align}
\mathbb{P}(S\neq\hat{S}')\leq \frac{H(S|\tilde{S})}{2\log 2}.\label{eq:HVineqnew}
\end{align}
It can be verified that
\begin{align}
\sum\limits_{s,\hat{s}\in\mathcal{S}:s\neq\hat{s}}p_{\hat{S}|S}(\hat{s}|s)&\leq\sum\limits_{s,\hat{s}\in\mathcal{S}:s\neq\hat{s}}p_{\hat{S}'|S}(\hat{s}|s)\nonumber\\
&\leq\frac{1}{\rho}\sum\limits_{s,\hat{s}\in\mathcal{S}:s\neq\hat{s}}p_S(s)p_{\hat{S}'|S}(\hat{s}|s)\nonumber\\
&=\frac{\mathbb{P}(S\neq\hat{S}')}{\rho}.\label{eq:offdiagnew}
\end{align}
Substituting (\ref{eq:HVineqnew}) into (\ref{eq:offdiagnew}) yields
\begin{align}
\sum\limits_{s,\hat{s}\in\mathcal{S}:s\neq\hat{s}}p_{\hat{S}|S}(\hat{s}|s)\leq\hbar\triangleq\frac{H(S|\tilde{S})}{2\rho\log 2}.\label{eq:tbinvokenew}
\end{align}
Note that
\begin{align}
\max\limits_{s\in\mathcal{S},\hat{s}\in\mathcal{S}_+:s\neq\hat{s}}\frac{p_{\hat{S}|S}(\hat{s}|s)}{\sum_{s'\in\mathcal{S}}p_{\hat{S}|S}(\hat{s}|s')}\leq\frac{\hbar}{\hbar+\mathbb{I}(\hbar\leq 1)}.\label{eq:twocases}
\end{align}
Indeed, (\ref{eq:twocases}) is trivially true when $\hbar>1$; moreover, when $\hbar\leq 1$,
\begin{align}
&\max\limits_{s\in\mathcal{S},\hat{s}\in\mathcal{S}_+:s\neq\hat{s}}\frac{p_{\hat{S}|S}(\hat{s}|s)}{\sum_{s'\in\mathcal{S}}p_{\hat{S}|S}(\hat{s}|s')}\nonumber\\
&\leq\max\limits_{s\in\mathcal{S},\hat{s}\in\mathcal{S}_+:s\neq\hat{s}}\frac{p_{\hat{S}|S}(\hat{s}|s)}{p_{\hat{S}|S}(\hat{s}|s)+p_{\hat{S}|S}(\hat{s}|\hat{s})}\nonumber\\
&=\max\limits_{s\in\mathcal{S},\hat{s}\in\mathcal{S}_+:s\neq\hat{s}}\frac{p_{\hat{S}|S}(\hat{s}|s)}{p_{\hat{S}|S}(\hat{s}|s)+1-\sum_{\hat{s}'\in\mathcal{S}:\hat{s}'\neq\hat{s}}p_{\hat{S}|S}(\hat{s}'|\hat{s})}\nonumber\\
&\leq\max\limits_{s\in\mathcal{S},\hat{s}\in\mathcal{S}_+:s\neq\hat{s}}\frac{p_{\hat{S}|S}(\hat{s}|s)}{2p_{\hat{S}|S}(\hat{s}|s)+1-\hbar}\label{eq:h1}\\
&\leq\frac{\hbar}{\hbar+1},\label{eq:h2}
\end{align}
where (\ref{eq:h1}) and (\ref{eq:h2}) are due to (\ref{eq:tbinvokenew}). In view of Theorem \ref{thm:theorem2variant},
It suffices to have
\begin{align}
\frac{\hbar}{\hbar+\mathbb{I}(\hbar\leq 1)}\leq\frac{1}{(|\mathcal{S}|-1)e-|\mathcal{S}|+2}.\label{eq:hbar}
\end{align}
Note that (\ref{eq:hbar}) is not satisfied when $\hbar>1$ since its left-hand side is equal to 1 whereas its right-hand side is strictly less than 1 ($\hbar>1$  implies $|\mathcal{S}|\geq 2$). When $\hbar\leq 1$, we can rewrite (\ref{eq:hbar}) as\footnote{Note that $\hbar\leq\frac{1}{(|\mathcal{S}|-1)(e-1)}$ implies $\hbar\leq 1$ when $|\mathcal{S}|\geq 2$. The case $|\mathcal{S}|=1$ is trivial since $\hbar$ can only take the value 0.}
\begin{align*}
\hbar\leq\frac{1}{(|\mathcal{S}|-1)(e-1)},
\end{align*}
which is exactly the desired result. This completes the proof of Theorem \ref{thm:theorem2}.

In Appendix \ref{app:alternative}, we give an alternative proof of Theorem \ref{thm:theorem2} with a different threshold on $H(S|\tilde{S})$.













\section{Extension and Discussion}\label{sec:discussion}

\subsection{Extension of Theorem \ref{thm:theorem1}}\label{subsec:extensionthm1}


It is interesting to know to what extent Theorem \ref{thm:theorem1} can be extended beyond the binary-input case. This subsection is largely devoted to answering this question. For any $p_{Y|X,S}$ and $p_S$, define
\begin{align*}
&\underline{\epsilon}(p_{Y|X,S},p_S)=\min\{\epsilon\in[0,1]:C(p_{Y|X,S},p_S,p_{\tilde{S}_{GE}^{(\epsilon)}|S})\\
&\hspace{2.1in}=\underline{C}(p_{Y|X,S},p_S)\},\\
&\underline{q}(p_{Y|X,S},p_S)=\min\{q\in[0,\frac{1}{|\mathcal{S}|}]:C(p_{Y|X,S},p_S,p_{\tilde{S}_{GS}^{(q)}|S})\\
&\hspace{2.25in}=\underline{C}(p_{Y|X,S},p_S)\}.
\end{align*}

\begin{proposition}\label{prop:ns1}
\begin{enumerate}
\item  There exists $\alpha(p_{Y|X,S},p_S)>0$ such that $C(p_{Y|X,S},p_S,p_{\tilde{S}|S})=\underline{C}(p_{Y|X,S},p_S)$
for all $p_{\tilde{S}|S}$ satisfying $I(S;\tilde{S})\leq\alpha(p_{Y|X,S},p_S)$ if and only if $\underline{\epsilon}(p_{Y|X,S},p_S)<1$.

\item $\underline{\epsilon}(p_{Y|X,S},p_S)<1$ if and only if
\begin{align}
&\sum\limits_{y\in\mathcal{Y},s\in\mathcal{S}} p_S(s)\delta(u,y,s)\nonumber\\
&\hspace{0.5in}\times\log\frac{p_{Y|X,S}(y|\psi(u,\ast),s)}{\sum_{x\in\mathcal{X}}p_{\hat{X}}(x)p_{Y|X,S}(y|x,s)}>0,\nonumber\\
&\hspace{2in} u\in\mathcal{U}_+\backslash\mathcal{G}_{\delta},\label{eq:thm1ns}
\end{align}
where $\delta(u,y,s)$ and $\mathcal{G}_{\delta}$ are defined in (\ref{eq:defdelta}) and (\ref{eq:defutilde}), respectively, $p_{\hat{X}}$ is an arbitrary maximizer of the optimization problem in (\ref{eq:nostate}), and
\begin{align*}
&\mathcal{U}_+=\Bigg\{u\in\mathcal{U}:\sum\limits_{y\in\mathcal{Y},s\in\mathcal{S}}p_S(s)p_{Y|X,S}(y|\psi(u,\ast),s)\nonumber\\
&\times\log\frac{p_{Y|X,S}(y|\psi(u,\ast),s)}{\sum_{x\in\mathcal{X}}p_{\hat{X}}(x)p_{Y|X,S}(y|x,s)}=\underline{C}(p_{Y|X,S},p_S)\Bigg\}.
\end{align*}
\end{enumerate}
\end{proposition}
\emph{Remark:} All maximizers of the optimization problem in (\ref{eq:nostate}) give rise to the same $\sum_{x\in\mathcal{X}}p_{\hat{X}}(x)p_{Y|X,S}(y|x,s)$, $y\in\mathcal{Y}$, $s\in\mathcal{S}$ \cite[p. 96, Cor. 2]{Gallager68}.
\begin{IEEEproof}
The first statement can be easily extracted from the proof of Theorem \ref{thm:theorem1}.

\begin{figure*}[tb]
\begin{centering}
\includegraphics[width=13cm]{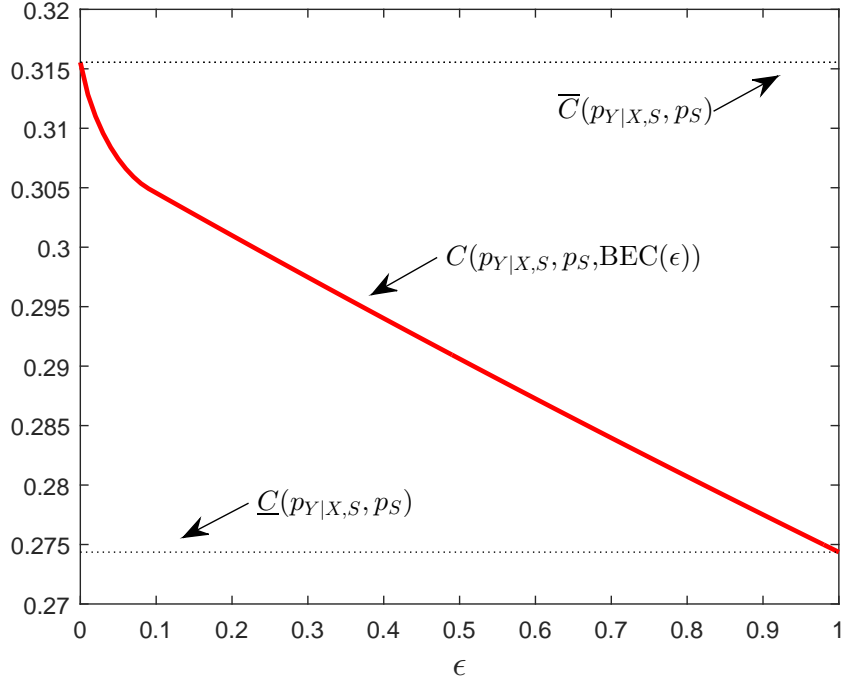}
\caption{Plot of $C(p_{Y|X,S},p_S,\mbox{BEC}(\epsilon))$ against $\epsilon$ for $\epsilon\in[0,1]$, where $p_{Y|X,S}$ and $p_S$ are given by (\ref{eq:counterthm1a}) and (\ref{eq:counterthm1b}), respectively. \label{fig:thm1nonbinary}}
\end{centering}
\end{figure*}

Now we proceed to prove the second statement. First recall the definitions of  $D_{GE}(p_U,\epsilon,u)$ and $p_{\hat{U}}$ in (\ref{eq:defD}) and (\ref{eq:uconstruction}), respectively. Since $p_{\hat{U}}$ is a capacity-achieving input distribution of channel $p_{Y,S|U}$ when $\epsilon=1$, we must have
\begin{align*}
&D_{GE}(p_{\hat{U}},1,u)=\underline{C}(p_{Y|X,S},p_S),\quad u\in\mathcal{U}\mbox{ with }p_{\hat{U}}(u)>0, \\
&D_{GE}(p_{\hat{U}},1,u)\leq \underline{C}(p_{Y|X,S},p_S),\quad u\in\mathcal{U}\mbox{ with }p_{\hat{U}}(u)=0,
\end{align*}
which, together with the fact $\mathcal{U}_+=\{u\in\mathcal{U}:D_{GE}(p_{\hat{U}},1,u)=\underline{C}(p_{Y|X,S},p_S)\}$, implies
\begin{align}
&\{u\in\mathcal{U}:p_{\hat{U}}(u)>0\}\subseteq\mathcal{U}_+,\label{eq:intersect1}\\
&D_{GE}(p_{\hat{U}},1,u)=\underline{C}(p_{Y|X,S},p_S),\quad u\in\mathcal{U}_+,\label{eq:xx1}\\
&D_{GE}(p_{\hat{U}},1,u)<\underline{C}(p_{Y|X,S},p_S),\quad \mbox{otherwise}.\label{eq:xx2}
\end{align}
It can be verified that
\begin{align}
&D_{GE}(p_{\hat{U}},\epsilon,u)=D_{GE}(p_{\hat{U}},1,u),\quad \epsilon\in[0,1], u\in\mathcal{G}_{\delta}.\label{eq:xx3}
\end{align}
Moreover, in view of (\ref{eq:newsub}), we can write (\ref{eq:thm1ns}) equivalently as
\begin{align}
\left.\frac{\partial}{\partial\epsilon}D_{GE}(p_{\hat{U}},\epsilon,u)\right|_{\epsilon=1}>0,\quad u\in\mathcal{U}_+\backslash\mathcal{G}_{\delta}.\label{eq:xx4}
\end{align}
According to (\ref{eq:xx1})--(\ref{eq:xx4}), there exists $\epsilon(p_{Y|X,S},p_S)\in[0,1)$ such that
\begin{align}
&D_{GE}(p_{\hat{U}},\epsilon,u)=\underline{C}(p_{Y|X,S},p_S),\quad u\in\mathcal{U}_+\cap\mathcal{G}_{\delta},\label{eq:KKT1}\\
&D_{GE}(p_{\hat{U}},\epsilon,u)\leq\underline{C}(p_{Y|X,S},p_S),\quad\mbox{otherwise}\label{eq:KKT2}
\end{align}
for $\epsilon\geq\epsilon(p_{Y|X,S},p_S)$. In light of (\ref{eq:intersect1}) and the fact $\{u\in\mathcal{U}:p_{\hat{U}}(u)>0\}\subseteq\mathcal{X}\subseteq\mathcal{G}_{\delta}$,
we have
\begin{align}
\{u\in\mathcal{U}:p_{\hat{U}}(u)>0\}\subseteq\mathcal{U}_+\cap\mathcal{G}_{\delta}.\label{eq:intersect2}
\end{align}
Combining (\ref{eq:KKT1}), (\ref{eq:KKT2}), and (\ref{eq:intersect2}) proves the ``if" part of the second statement.
Next we turn to the ``only if" part of the second statement. Assuming the existence of $\epsilon(p_{Y|X,S},p_S)\in[0,1)$ such that $C(p_{Y|X,S},p_S,p_{\tilde{S}^{(\epsilon)}|S})=\underline{C}(p_{Y|X,S},p_S)$ for $\epsilon\geq\epsilon(p_{Y|X,S},p_S)$ (or equivalently $p_{\hat{U}}$ is a capacity-achieving input distribution of channel $p_{Y,S|U}$ for $\epsilon\geq\epsilon(p_{Y|X,S},p_S)$), we must have
\begin{align}
D_{GE}(p_{\hat{U}},\epsilon,u)\leq\underline{C}(p_{Y|X,S},p_S),\quad \epsilon\geq\epsilon(p_{Y|X,S},p_S), u\in\mathcal{U}.\label{eq:lower}
\end{align}
It can be verified that
\begin{align}
&\frac{\partial^2}{\partial\epsilon^2}D_{GE}(p_{\hat{U}},\epsilon,u)\nonumber\\
&=\sum\limits_{y\in\mathcal{Y},s\in\mathcal{S}}\frac{p_S(s)\delta^2(u,y,s)}{p_{Y|X,S}(y|\psi(u,s),s)+\epsilon\delta(u,y,s)}\nonumber\\
&>0,\quad \epsilon\in[0,1], u\in\mathcal{U}\backslash\mathcal{G}_{\delta}.\label{eq:secondderivative}
\end{align}
Moreover, by the definition of $\mathcal{U}_+$,
\begin{align}
D_{GE}(p_{\hat{U}},1,u)=\underline{C}(p_{Y|X,S},p_S),\quad u\in\mathcal{U}_+.\label{eq:boundary}
\end{align}
Note that (\ref{eq:lower}), (\ref{eq:secondderivative}), and (\ref{eq:boundary})  hold simultaneously for $u\in\mathcal{U}_+\backslash\mathcal{G}_{\delta}$, from which (\ref{eq:thm1ns}) (or equivalently (\ref{eq:xx4})) can be readily deduced. This completes the proof of Proposition \ref{prop:ns1}.
\end{IEEEproof}

As shown by the following example, the necessary and sufficient condition (\ref{eq:thm1ns}) is not always satisfied when $|\mathcal{X}|>2$. Let
\begin{align}
&p_{Y|X,S}(y|x,s)=\left\{
                   \begin{array}{ll}
                     1, & (x,y,s)=(0,0,0)\mbox{ or }(1,1,1), \\
                     0, & (x,y,s)=(0,1,0)\mbox{ or }(1,0,1), \\
                     \frac{2}{5}, & (x,y,s)=(1,0,0)\mbox{ or }(0,1,1), \\
                     \frac{3}{5}, & (x,y,s)=(1,1,0)\mbox{ or }(0,0,1), \\
                     \frac{3}{10}, & (x,y,s)=(2,0,0), \\
                     \frac{1}{5}, & (x,y,s)=(2,0,1),\\
                     \frac{7}{10}, & (x,y,s)=(2,1,0),\\
                      \frac{4}{5}, & (x,y,s)=(2,1,1),
                   \end{array}
                 \right.\label{eq:counterthm1a}\\
&p_S(0)=p_S(1)=\frac{1}{2}.\label{eq:counterthm1b}
\end{align}
For this example, it can be verified that $\hat{u}\in\mathcal{U}_+\backslash\mathcal{G}_{\delta}$ and
\begin{align*}
\sum\limits_{y\in\mathcal{Y},s\in\mathcal{S}} p_S(s)\delta(\hat{u},y,s)\log\frac{p_{Y|X,S}(y|\psi(\hat{u},\ast),s)}{\sum_{x\in\mathcal{X}}p_{\hat{X}}(x)p_{Y|X,S}(y|x,s)}<0,
\end{align*}
where $\psi(\hat{u},\cdot)$ is given by $\psi(\hat{u},0)=2$, $\psi(\hat{u},1)=1$, and $\psi(\hat{u},\ast)=1$; indeed, Fig. \ref{fig:thm1nonbinary} shows that $C(p_{Y|X,S},p_S,\mbox{BEC}(\epsilon))>\underline{C}(p_{Y|X,S},p_S)$ for $\epsilon\in[0,1)$.

The proof of Proposition \ref{prop:ns1} in fact suggests a strategy for computing $\underline{\epsilon}(p_{Y|X,S},p_S)$. Let $p_{\hat{X}}$ be an arbitrary maximizer of the optimization problem in (\ref{eq:nostate}) and define $p_{\hat{U}}$ according to (\ref{eq:uconstruction}). Note that
\begin{itemize}
\item $D_{GE}(p_{\hat{U}},1,u)\leq\underline{C}(p_{Y|X,S},p_S)$ for $u\in\mathcal{U}$ (see (\ref{eq:xx1}) and (\ref{eq:xx2})),

\item $D_{GE}(p_{\hat{U}},\epsilon,u)$ does not depend on $\epsilon$ for $u\in\mathcal{G}_{\delta}$ (see (\ref{eq:xx3})),

\item $D_{GE}(p_{\hat{U}},\epsilon,u)$ is a strictly convex function of $\epsilon$ for $u\in\mathcal{U}\backslash\mathcal{G}_{\delta}$ (see (\ref{eq:secondderivative})).
\end{itemize}
Hence, for each $u\in\mathcal{U}$, there are three mutually exclusive cases.
\begin{enumerate}
\item $D_{GE}(p_{\hat{U}},0,u)\leq\underline{C}(p_{Y|X,S},p_S)$: We have $D_{GE}(p_{\hat{U}},\epsilon,u)\leq\underline{C}(p_{Y|X,S},p_S)$ for $\epsilon\in[\epsilon(u),1]$, where $\epsilon(u)=0$.

\item $D_{GE}(p_{\hat{U}},0,u)>D_{GE}(p_{\hat{U}},1,u)=\underline{C}(p_{Y|X,S},p_S)$ and $\left.\frac{\partial}{\partial\epsilon}D_{GE}(p_{\hat{U}},\epsilon,u)\right|_{\epsilon=1}\leq 0$ (this case can arise only when $|\mathcal{X}|>2$): We have $D_{GE}(p_{\hat{U}},0,u)>\underline{C}(p_{Y|X,S},p_S)$ for $\epsilon\in[0,\epsilon(u))$, where $\epsilon(u)=1$.

\item Otherwise: We have $D_{GE}(p_{\hat{U}},\epsilon,u)>\underline{C}(p_{Y|X,S},p_S)$ for $\epsilon\in[0,\epsilon(u))$ and $D_{GE}(p_{\hat{U}},\epsilon,u)\leq\underline{C}(p_{Y|X,S},p_S)$ for $\epsilon\in[\epsilon(u),1]$, where $\epsilon(u)$ is  the unique solution of $D_{GE}(p_{\hat{U}},\epsilon,u)=\underline{C}(p_{Y|X,S},p_S)$ for $\epsilon\in(0,1)$.
\end{enumerate}
It can be readily shown that
\begin{align}
\underline{\epsilon}(p_{Y|X,S},p_S)=\max\limits_{u\in\mathcal{U}}\epsilon(u).\label{eq:maxepsilon}
\end{align}



We can compute $\underline{q}(p_{Y|X,S},p_S)$ in a similar way. Define
\begin{align*}
D_{GS}(p_U,q,u)=D(p_{Y,S|U}(\cdot,\cdot|u)\|p_{Y,S}),
\end{align*}
where
\begin{align*}
p_{Y,S|U}(y,s|u)&=p_S(s)(p_{Y|X,S}(y|\psi(u,s),s)+q\omega(u,y,s))
\end{align*}
with
\begin{align*}
\omega(u,y,s)&=\sum\limits_{\tilde{s}\in\mathcal{S}:\tilde{s}\neq s}p_{Y|X,S}(y|\psi(u,\tilde{s}),s)\\
&\quad-(|\mathcal{S}|-1)p_{Y|X,S}(y|\psi(u,s),s),\\
&\hspace{1.5in} u\in\mathcal{U},y\in\mathcal{Y},s\in\mathcal{S}.
\end{align*}
Again, let $p_{\hat{U}}$ be defined\footnote{Note that the underlying $\mathcal{U}$ depends on $\tilde{\mathcal{S}}$. In particular, $|\mathcal{U}|=|\mathcal{X}|^{|\mathcal{S}|}$ when $p_{\tilde{S}|S}$ is a generalized symmetric channel whereas $|\mathcal{U}|=|\mathcal{X}|^{|\mathcal{S}|+1}$ when $p_{\tilde{S}|S}$ is a generalized erasure channel.} according to (\ref{eq:uconstruction}).
It can be verified that
\begin{align*}
&D_{GS}(p_{\hat{U}},q,u)\\
&=\sum\limits_{y\in\mathcal{Y},s\in\mathcal{S}} p_S(s)(p_{Y|X,S}(y|\psi(u,s),s)+q\omega(u,y,s))\\
&\hspace{0.5in}\times\log\frac{p_{Y|X,S}(y|\psi(u,s),s)+q\omega(u,y,s)}{\sum_{x\in\mathcal{X}}p_{\hat{X}}(x)p_{Y|X,S}(y|x,s)},\\
&\hspace{2in} q\in[0,\frac{1}{|\mathcal{S}|}], u\in\mathcal{U},\\
&\frac{\partial}{\partial q}D_{GS}(p_{\hat{U}},q,u)\\
&=\sum\limits_{y\in\mathcal{Y},s\in\mathcal{S}} p_S(s)\delta(u,y,s)\\
&\hspace{0.5in}\times\log\frac{p_{Y|X,S}(y|\psi(u,s),s)+q\omega(u,y,s)}{\sum_{x\in\mathcal{X}}p_{\hat{X}}(x)p_{Y|X,S}(y|x,s)},\\
&\hspace{2in} q\in[0,\frac{1}{|\mathcal{S}|}], u\in\mathcal{U},\\
&\frac{\partial^2}{\partial q^2}D_{GS}(p_{\hat{U}},q,u)\\
&=\sum\limits_{y\in\mathcal{Y},s\in\mathcal{S}}\frac{p_S(s)\delta^2(u,y,s)}{p_{Y|X,S}(y|\psi(u,s),s)+q\omega(u,y,s)}>0,\\
&\hspace{2in}q\in[0,\frac{1}{|\mathcal{S}|}], u\in\mathcal{U}\backslash\mathcal{G}_{\omega},
\end{align*}
where
\begin{align*}
\mathcal{G}_{\omega}=\{u\in\mathcal{U}: \omega(u,y,s)=0\mbox{ for all }y\in\mathcal{Y}\mbox{ and }s\in\mathcal{S}\}.
\end{align*}
Clearly,
\begin{itemize}
\item $D_{GS}(p_{\hat{U}},\frac{1}{|\mathcal{S}|},u)\leq\underline{C}(p_{Y|X,S},p_S)$ for $u\in\mathcal{U}$,

\item $D_{GS}(p_{\hat{U}},q,u)$ does not depend on $q$ for $u\in\mathcal{G}_{\omega}$,

\item $D_{GS}(p_{\hat{U}},q,u)$ is a strictly convex function of $q$ for $u\in\mathcal{U}\backslash\mathcal{G}_{\omega}$.
\end{itemize}
Hence, for each $u\in\mathcal{U}$, there are also three mutually exclusive cases.
\begin{enumerate}
\item $D_{GS}(p_{\hat{U}},0,u)\leq\underline{C}(p_{Y|X,S},p_S)$: We have $D_{GS}(p_{\hat{U}},q,u)\leq\underline{C}(p_{Y|X,S},p_S)$ for $q\in[q(u),1]$, where $q(u)=0$.

\item $D_{GS}(p_{\hat{U}},0,u)>D_{GS}(p_{\hat{U}},\frac{1}{|\mathcal{S}|},u)=\underline{C}(p_{Y|X,S},p_S)$ and $\left.\frac{\partial}{\partial q}D_{GS}(p_{\hat{U}},q,u)\right|_{q=\frac{1}{|\mathcal{S}|}}\leq 0$ (this case can arise only when $|\mathcal{X}|>2$): We have $D_{GS}(p_{\hat{U}},0,u)>\underline{C}(p_{Y|X,S},p_S)$ for $q\in[0,q(u))$, where $q(u)=\frac{1}{|\mathcal{S}|}$.

\item Otherwise: We have $D_{GS}(p_{\hat{U}},q,u)>\underline{C}(p_{Y|X,S},p_S)$ for $q\in[0,q(u))$ and $D_{GS}(p_{\hat{U}},q,u)\leq\underline{C}(p_{Y|X,S},p_S)$ for $q\in[q(u),\frac{1}{|\mathcal{S}|}]$, where $q(u)$ is  the unique solution of $D_{GS}(p_{\hat{U}},q,u)=\underline{C}(p_{Y|X,S},p_S)$ for $q\in(0,\frac{1}{|\mathcal{S}|})$.
\end{enumerate}
It can be readily shown that
\begin{align}
\underline{q}(p_{Y|X,S},p_S)=\max\limits_{u\in\mathcal{U}}q(u).\label{eq:maxq}
\end{align}

For $p_{Y|X,S}$ and $p_S$ illustrated in Fig. \ref{fig:SZchannel} (see also (\ref{eq:para1}) and (\ref{eq:para2})), we show in Appendix \ref{app:underline} that
\begin{align}
&\underline{\epsilon}(p_{Y|X,S},p_S)=\left\{
                                      \begin{array}{ll}
\hat{\epsilon}(\theta), & \theta\in(0,1),\\
                                        0, & \mbox{otherwise},
                                      \end{array}
                                    \right.\label{eq:proofinAppB1}\\
&\underline{q}(p_{Y|X,S},p_S)=\left\{
                                      \begin{array}{ll}
\hat{q}(\theta), & \theta\in(0,1),\\
                                        0, & \mbox{otherwise},
                                      \end{array}
                                    \right.\label{eq:proofinAppB2}
\end{align}
where $\hat{\epsilon}(\theta)$ is the unique solution of
\begin{align*}
&\epsilon(1-\theta)\log 2\epsilon+(1-\epsilon(1-\theta))\log\frac{2(1-\epsilon(1-\theta))}{1+\theta}\\
&=(1-\theta)\log2+\theta\log\frac{2\theta}{1+\theta}
\end{align*}
for $\epsilon\in(0,1)$, and $\hat{q}(\theta)$ is the unique solution of
\begin{align*}
&q(1-\theta)\log2q+(1-q(1-\theta))\log\frac{2(1-q(1-\theta))}{1+\theta}\\
&=\frac{1}{2}\Bigg((1-\theta)\log2+\log\frac{2}{1+\theta}+\theta\log\frac{2\theta}{1+\theta}\Bigg)
\end{align*}
for $q\in(0,\frac{1}{2})$. Setting $\theta=\frac{1}{2}$ gives $\underline{\epsilon}(p_{Y|X,S},p_S)\approx0.1$ (cf. Fig. \ref{fig:plot1}) and $\underline{q}(p_{Y|X,S},p_S)\approx0.037$ (cf. Fig. \ref{fig:plot2}).



\subsection{Extension of Theorem \ref{thm:theorem2}}

\begin{figure*}[tb]
\begin{centering}
\includegraphics[width=13cm]{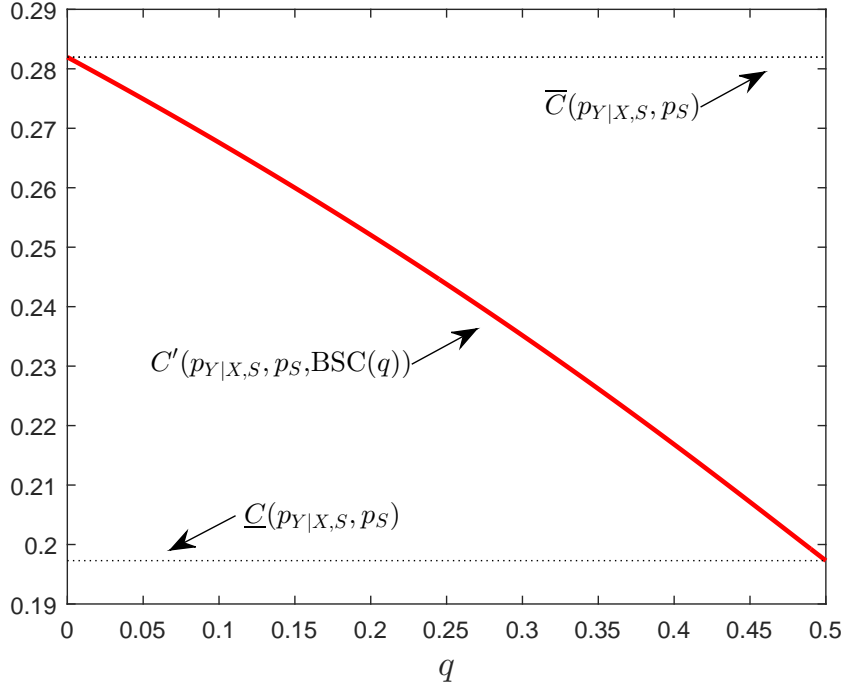}
\caption{Plot of $C'(p_{Y|X,S},p_S,\mbox{BSC}(q))$ against $q$ for $q\in[0,\frac{1}{2}]$, where  $p_{Y|X,S}$ and $p_S$ are given by (\ref{eq:counterthm2a}) and (\ref{eq:counterthm2b}), respectively. \label{fig:thm2nonbinary}}
\end{centering}
\end{figure*}

We shall extend Theorem \ref{thm:theorem2} in a similar fashion. For any $p_{Y|X,S}$ and $p_S$, define
\begin{align*}
&\overline{\epsilon}(p_{Y|X,S},p_S)=\max\{\epsilon\in[0,1]:C'(p_{Y|X,S},p_S,p_{\tilde{S}_{GE}^{(\epsilon)}|S})\\
&\hspace{2.2in}=\overline{C}(p_{Y|X,S},p_S)\},\\
&\overline{q}(p_{Y|X,S},p_S)=\max\{q\in[0,\frac{1}{|\mathcal{S}|}]:C'(p_{Y|X,S},p_S,p_{\tilde{S}_{GS}^{(q)}|S})\\
&\hspace{2.3in}=\overline{C}(p_{Y|X,S},p_S)\}.
\end{align*}

\begin{proposition}\label{prop:ns2}
\begin{enumerate}
\item There exists $\beta(p_{Y|X,S},p_S)>0$ such that $C'(p_{Y|X,S},p_S,p_{\tilde{S}|S})=\overline{C}(p_{Y|X,S},p_S)$
for all $p_{\tilde{S}|S}$ satisfying $H(S|\tilde{S})\leq\beta(p_{Y|X,S},p_S)$ if and only if $\overline{q}(p_{Y|X,S},p_S)>0$.

\item $\overline{q}(p_{Y|X,S},p_S)>0$ if and only if there exists $p_{\hat{X}|S}\in\mathcal{P}$ such that
\begin{align}
\{x\in\mathcal{X}:p_{\hat{X}|S}(x|s)>0\}=\mathcal{X}_+,\quad s\in\mathcal{S}.\label{eq:constantset}
\end{align}
\end{enumerate}
\end{proposition}
\begin{IEEEproof}
The first statement can be easily extracted from the proof of Theorem \ref{thm:theorem2}. The second statement is a consequence of Lemma \ref{lem:nssym}.
\end{IEEEproof}

As shown by the following example, the necessary and sufficient condition (\ref{eq:constantset}) is not always satisfied when $|\mathcal{X}|>2$. Let
\begin{align}
&p_{Y|X,S}(y|x,s)=\left\{
                   \begin{array}{ll}
                     1, & (x,y,s)=(0,0,0)\mbox{ or }(2,1,1), \\
                     0, & (x,y,s)=(0,1,0)\mbox{ or }(2,0,1), \\
                     \frac{2}{5}, & (x,y,s)=(1,0,0)\mbox{ or }(0,1,1), \\
                     \frac{3}{5}, & (x,y,s)=(1,1,0)\mbox{ or }(0,0,1), \\
                     \frac{4}{5}, & (x,y,s)=(2,0,0)\mbox{ or }(1,1,1), \\
                     \frac{1}{5}, & (x,y,s)=(2,1,0)\mbox{ or }(1,0,1),
                   \end{array}
                 \right.\label{eq:counterthm2a}\\
&p_S(0)=p_S(1)=\frac{1}{2}.\label{eq:counterthm2b}
\end{align}
For this example, it can be verified that the maximizer of the optimization problem in (\ref{eq:perfectstate}), denoted by $p_{\hat{X}|S}$, is unique and
\begin{align*}
&\{x\in\mathcal{X}:p_{\hat{X}|S}(x|0)>0\}=\{0,1\},\\
&\{x\in\mathcal{X}:p_{\hat{X}|S}(x|1)>0\}=\{0,2\};
\end{align*}
indeed, Fig. \ref{fig:thm2nonbinary} shows that $C'(p_{Y|X,S},p_S,\mbox{BSC}(q))<\overline{C}(p_{Y|X,S},p_S)$ for $q\in(0,\frac{1}{2}]$.

In view of Lemmas \ref{lem:erasure} and \ref{lem:nssym}, we have
\begin{align}
&\overline{\epsilon}(p_{Y|X,S},p_S)=\max\limits_{p_{\hat{X}|S}\in\mathcal{P}}\sum\limits_{x\in\mathcal{X}}\min\limits_{s\in\mathcal{S}}p_{\hat{X}|S}(x|s),\label{eq:invewepsilon}\\
&\overline{q}(p_{Y|X,S},p_S)=\max\limits_{p_{\hat{X}|S}\in\mathcal{P}}\min\limits_{x\in\mathcal{X}_+,s\in\mathcal{S}}\frac{p_{\hat{X}|S}(x|s)}{\sum_{s'\in\mathcal{S}}p_{\hat{X}|S}(x|s')}.\label{eq:inviewq}
\end{align}
Note that $\mathcal{P}$ does not depend on $p_S$ (under the assumption $\rho>0$); as a consequence, $\overline{\epsilon}(p_{Y|X,S},p_S)$ and $\overline{q}(p_{Y|X,S},p_S)$ do not depend on $p_S$ either. For $p_{Y|X,S}$ and $p_S$ illustrated in Fig. \ref{fig:SZchannel} (see also (\ref{eq:para1}) and (\ref{eq:para2})), we show in Appendix \ref{app:overline} that
\begin{align}
&\overline{\epsilon}(p_{Y|X,S},p_S)\nonumber\\
&=\left\{
                                      \begin{array}{ll}
2\Bigg(1+(1-\theta)\theta^{\frac{\theta}{1-\theta}}\Bigg)^{-1}\theta^{\frac{\theta}{1-\theta}}, & \theta\in(0,1),\\
                                        1, & \mbox{otherwise},
                                      \end{array}
                                    \right.\label{eq:proofinAppC1}\\
&\overline{q}(p_{Y|X,S},p_S)\nonumber\\
&=\left\{
                               \begin{array}{ll}
                                 \Bigg(1+(1-\theta)\theta^{\frac{\theta}{1-\theta}}\Bigg)^{-1}\theta^{\frac{\theta}{1-\theta}}, &\theta\in(0,1), \\
                                 \frac{1}{2}, & \mbox{otherwise}.
                               \end{array}
                             \right.\label{eq:proofinAppC2}
\end{align}
Setting $\theta=\frac{1}{2}$ gives $\overline{\epsilon}(p_{Y|X,S},p_S)=\frac{4}{5}$ (cf. Fig. \ref{fig:plot1}) and $\overline{q}(p_{Y|X,S},p_S)=\frac{2}{5}$ (cf. Fig. \ref{fig:plot2}).

\subsection{Two Implicit Conditions}

\begin{figure*}[tb]
\begin{centering}
\includegraphics[width=13cm]{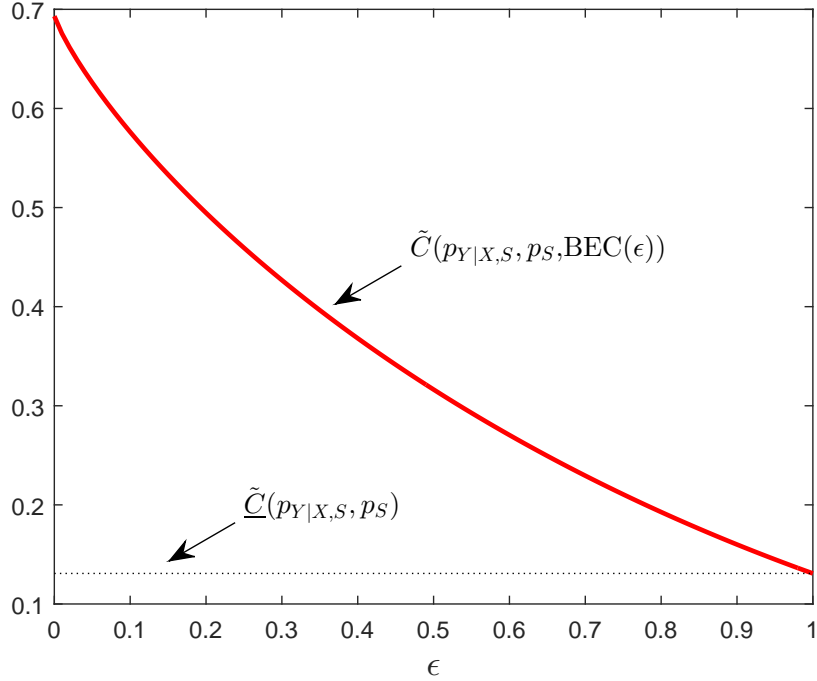}
\caption{Plot of $\tilde{C}(p_{Y|X,S},p_S,\mbox{BEC}(\epsilon))$ against $\epsilon$ for $\epsilon\in[0,1]$, where $p_{Y|X,S}$ and $p_S$ are given by (\ref{eq:channel}) with $\mu=\frac{1}{4}$ and (\ref{eq:state}), respectively.  \label{fig:nsdecoder}}
\end{centering}
\end{figure*}

In this subsection, we shall examine the following two implicit conditions in Theorem \ref{thm:theorem1}:
\begin{enumerate}
\item perfect state information at the decoder,

\item causal noisy state observation at the encoder.
\end{enumerate}

\begin{figure*}[tb]
\begin{centering}
\includegraphics[width=13cm]{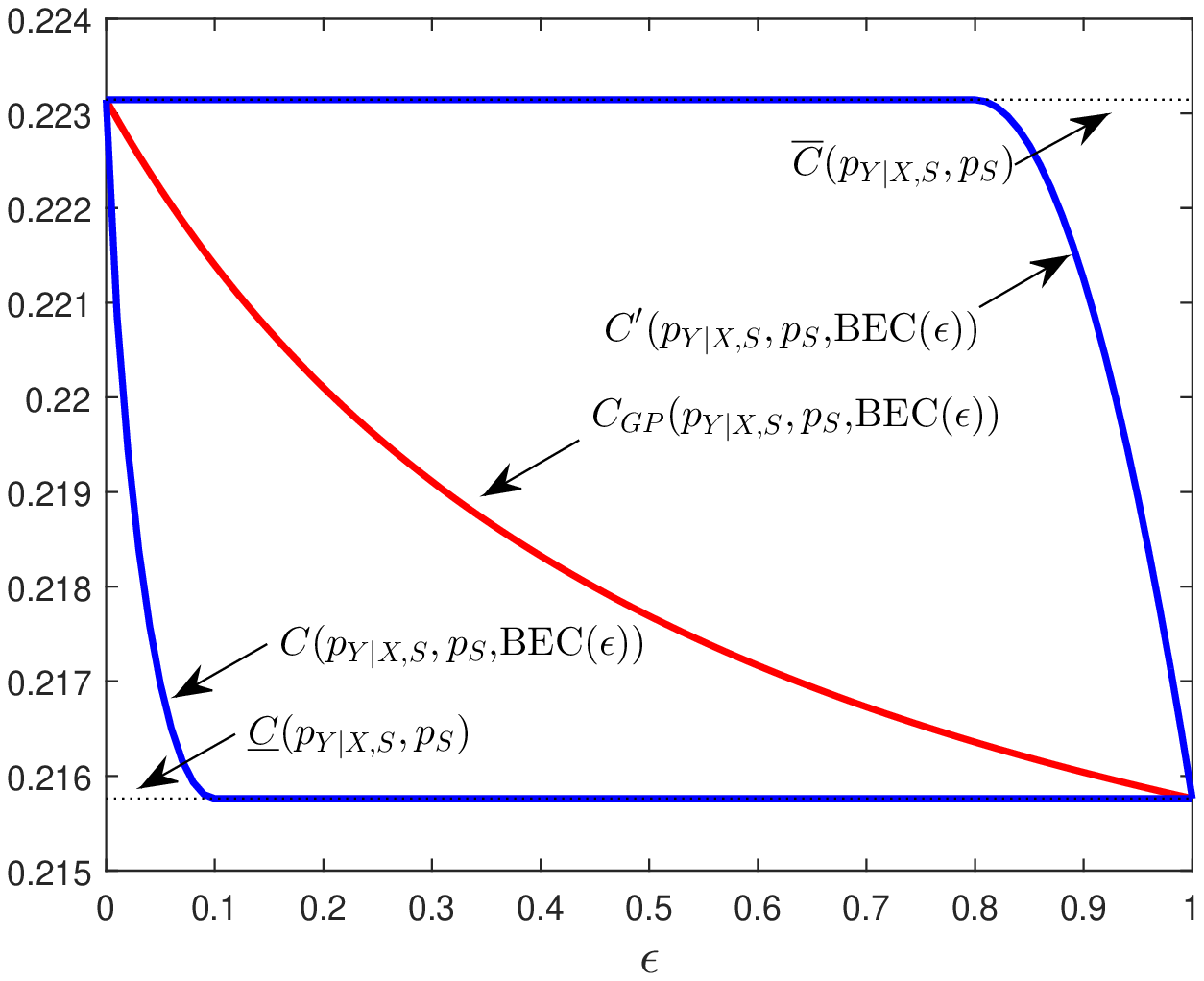}
\caption{Plot of $C_{GP}(p_{Y|X,S},p_S,\mbox{BEC}(\epsilon))$ against $\epsilon$ for $\epsilon\in[0,1]$, where  $p_{Y|X,S}$ and $p_S$ are given by (\ref{eq:para1}) with $\theta=\frac{1}{2}$ and (\ref{eq:para2}), respectively. \label{fig:GP}}
\end{centering}
\end{figure*}

If no state information is available at the decoder, then the channel capacity is given by
\begin{align*}
\tilde{C}(p_{Y|X,S},p_S,p_{\tilde{S}|S})\triangleq\max\limits_{p_U}I(U;Y),
\end{align*}
where the joint distribution of $(U,X,Y,S,\tilde{S})$ is given by (\ref{eq:jointdistribution}).
Furthermore, if there is also no state information available at the encoder, then the channel capacity becomes
\begin{align}
\underline{\tilde{C}}(p_{Y|X,S},p_S)\triangleq\max\limits_{p_X}I(X;Y),\label{eq:noedstate}
\end{align}
where $p_{X,Y,S}(x,y,s)= p_{X}(x)p_S(s)p_{Y|X,S}(y|x,s)$. Define
\begin{align*}
&\underline{\tilde{\epsilon}}(p_{Y|X,S},p_S)=\min\{\epsilon\in[0,1]:\tilde{C}(p_{Y|X,S},p_S,p_{\tilde{S}_{GE}^{(\epsilon)}|S})\\
&\hspace{2.1in}=\underline{\tilde{C}}(p_{Y|X,S},p_S)\}.
\end{align*}
The proof of the following result is similar to that of Proposition \ref{prop:ns1} and is omitted.

\begin{proposition}\label{prop:ns3}
\begin{enumerate}
\item  There exists $\tilde{\alpha}(p_{Y|X,S},p_S)>0$ such that $\tilde{C}(p_{Y|X,S},p_S,p_{\tilde{S}|S})=\underline{\tilde{C}}(p_{Y|X,S},p_S)$
for all $p_{\tilde{S}|S}$ satisfying $I(S;\tilde{S})\leq\tilde{\alpha}(p_{Y|X,S},p_S)$ if and only if $\underline{\tilde{\epsilon}}(p_{Y|X,S},p_S)<1$.

\item $\underline{\tilde{\epsilon}}(p_{Y|X,S},p_S)<1$ if and only if
\begin{align}
&\sum\limits_{y\in\mathcal{Y}}\Bigg( \sum\limits_{s\in\mathcal{S}}p_S(s)\delta(u,y,s)\Bigg)\nonumber\\
&\times\log\frac{\sum_{s\in\mathcal{S}}p_S(s)p_{Y|X,S}(y|\psi(u,\ast),s)}{\sum_{x\in\mathcal{X},s\in\mathcal{S}}p_{\hat{X}}(x)p_S(s)p_{Y|X,S}(y|x,s)}>0,\nonumber\\
&\hspace{2in} u\in\tilde{\mathcal{U}}_+\backslash\tilde{\mathcal{G}}_{\delta},\label{eq:prop3ns}
\end{align}
where $\delta(u,y,s)$ is defined in (\ref{eq:defdelta}), $p_{\hat{X}}$ is an arbitrary maximizer of the optimization problem in (\ref{eq:noedstate}), and
\begin{align*}
&\tilde{\mathcal{G}}_{\delta}=\Bigg\{u\in\mathcal{U}: \sum\limits_{s\in\mathcal{S}}p_S(s)\delta(u,y,s)=0\mbox{ for all }y\in\mathcal{Y}\Bigg\},\\
&\tilde{\mathcal{U}}_+=\Bigg\{u\in\mathcal{U}:\sum\limits_{y\in\mathcal{Y}}\Bigg(\sum\limits_{s\in\mathcal{S}}p_S(s)p_{Y|X,S}(y|\psi(u,\ast),s)\Bigg)\\
&\hspace{0.5in}\times\log\frac{\sum_{s\in\mathcal{S}}p_S(s)p_{Y|X,S}(y|\psi(u,\ast),s)}{\sum_{x\in\mathcal{X},s\in\mathcal{S}}p_{\hat{X}}(x)p_{S}(s)p_{Y|X,S}(y|x,s)}\\
&\hspace{2.0in}=\underline{\tilde{C}}(p_{Y|X,S},p_S)\Bigg\}.
\end{align*}
\end{enumerate}
\end{proposition}

As shown by the following example, the necessary and sufficient condition (\ref{eq:prop3ns}) is not always satisfied even when $|\mathcal{X}|=2$. Let
\begin{align}
&Y=X\oplus S,\quad \mathcal{X}=\mathcal{Y}=\mathcal{S}=\{0,1\},\label{eq:channel}\\
&p_S(1)=\mu\in(0,\frac{1}{2}),\label{eq:state}
\end{align}
where $\oplus$ is the modulo-2 addition. It can be verified that (\ref{eq:prop3ns}) is not satisfied for this example; indeed, Fig. \ref{fig:nsdecoder} indicates that
\begin{align}
\tilde{C}(p_{Y|X,S},p_S,\mbox{BEC}(\epsilon))>\underline{\tilde{C}}(p_{Y|X,S},p_S),\quad\epsilon\in[0,1).\label{eq:example}
\end{align}
Here we give an alternative way to prove (\ref{eq:example}). Write $S=\tilde{S}\oplus\Delta$, where $\tilde{S}$ and $\Delta$ are two mutually independent Bernoulli random variables with
\begin{align*}
&p_{\tilde{S}}(1)=\nu\in[0,\mu],\\
&p_{\Delta}(1)=\frac{\mu-\nu}{1-2\nu}.
\end{align*}
It is clear that
\begin{align}
\tilde{C}(p_{Y|X,S},p_S,p_{\tilde{S}|S})&=\log 2-H(\Delta)\nonumber\\
&>\log 2-H(S)\nonumber\\
&=\underline{\tilde{C}}(p_{Y|X,S},p_S),\quad\nu\in(0,\mu].\label{eq:ex1}
\end{align}
In light of Lemma \ref{lem:erasure}, $p_{\tilde{S}|S}$  is a stochastically degraded version of $\mbox{BEC}(\epsilon)$ and consequently
\begin{align}
\tilde{C}(p_{Y|X,S},p_S,\mbox{BEC}(\epsilon))\geq\tilde{C}(p_{Y|X,S},p_S,p_{\tilde{S}|S})\label{eq:ex2}
\end{align}
if $H(S)-H(\Delta)\leq\frac{\mu^2(1-\epsilon)^2}{2}$. Combining (\ref{eq:ex1}) and (\ref{eq:ex2}) proves (\ref{eq:example}).






Now we proceed to examine the second implicit condition. If the noisy state observation is available non-causally at the encoder, the Gelfand-Pinsker theorem \cite{GP80} (see also \cite[Th. 7.3]{EGK11}) indicates that the channel capacity is given by
\begin{align*}
C_{GP}(p_{Y|X,S},p_S,p_{\tilde{S}|S})\triangleq\max\limits_{p_{U|\tilde{S}}}I(U;Y,S)-I(U;\tilde{S}),
\end{align*}
where the joint distribution of $(U,X,Y,S,\tilde{S})$ factors as
\begin{align*}
&p_{U,X,Y,S,\tilde{S}}(u,x,y,s,\tilde{s})\\
&= p_S(s)p_{\tilde{S}|S}(\tilde{s}|s)p_{U|\tilde{S}}(u|\tilde{s})\mathbb{I}(x=\psi(u,\tilde{s}))p_{Y|X,S}(y|x,s),\nonumber\\
&\hspace{1in}u\in\mathcal{U}, x\in\mathcal{X}, y\in\mathcal{Y}, s\in\mathcal{S}, \tilde{s}\in\tilde{S}.\nonumber
\end{align*}
It turns out that $C_{GP}(p_{Y|X,S},p_S,p_{\tilde{S}|S})$ is bounded between $C(p_{Y|X,S},p_S,p_{\tilde{S}|S})$ and $C'(p_{Y|X,S},p_S,p_{\tilde{S}|S})$, i.e.,
\begin{align*}
C(p_{Y|X,S},p_S,p_{\tilde{S}|S})&\leq C_{GP}(p_{Y|X,S},p_S,p_{\tilde{S}|S})\\
&\leq C'(p_{Y|X,S},p_S,p_{\tilde{S}|S}).
\end{align*}
Indeed, the first inequality is obvious, and the second one holds because
\begin{align*}
I(U;Y,S)-I(U;\tilde{S})&\leq I(U;Y,S)-I(U;S)\\
&=I(U;Y|S)\\
&\leq I(X;Y|S).
\end{align*}
In Fig. \ref{fig:GP} we plot $C_{GP}(p_{Y|X,S},p_S,\mbox{BEC}(\epsilon))$ against $\epsilon$ for $\epsilon\in[0,1]$, where $p_{Y|X,S}$ and $p_S$ are given by (\ref{eq:para1}) with $\theta=\frac{1}{2}$ and (\ref{eq:para2}), respectively; it can be seen that $C_{GP}(p_{Y|X,S},p_S,\mbox{BEC}(\epsilon))$ is strictly greater than $\underline{C}(p_{Y|X,S},p_S)$ except when $\epsilon=1$.
So the causality condition on the noisy state observation at the encoder is not superfluous for Theorem \ref{thm:theorem1}.

\section{Conclusion}\label{sec:conclusion}

We have shown that the capacity of binary-input\footnote{In fact, both numerical simulation and theoretical analysis suggest that similar results hold for many (but not all) non-binary input channels.} channels is very ``sensitive" to the quality of the encoder side information whereas the generalized probing capacity is very ``robust". Here the words ``sensitive" and ``robust" should not be understood in a quantitative sense. Indeed, it is known \cite{SF04} that, when $|\mathcal{X}|=2$, the ratio of $\underline{C}(p_{Y|X,S},p_S)$ to $\overline{C}(p_{Y|X,S},p_S)$ is at least 0.942 and the difference between these two quantities is at most $\sim$0.011 bit; in other words, the gain that can be obtained by exploiting the encoder side information (or the loss that can be incurred by ignoring the encoder side information) is very limited anyway.

Binary signalling is widely used, especially in wideband communications. So our work might have some practical relevance. However,  great caution should be exercised in interpreting Theorems \ref{thm:theorem1} and \ref{thm:theorem2}. Specifically, both results rely on the assumption that the channel state takes values from a finite set\footnote{In contrast, the assumption  $|\mathcal{Y}|<\infty$ and $|\tilde{\mathcal{S}}|<\infty$ is not essential}, which is not necessarily satisfied in reality; moreover, the freedom of power control in real communication systems is not captured by our results. Nevertheless, our work can be viewed as an initial step towards a better understanding of the fundamental performance limits of communication systems where the transmitter side information and the receiver side information are not deterministically related.

Finally, it is worth mentioning that our results might have their counterparts in source coding.






\appendices

\section{An Alternative Proof of Theorem \ref{thm:theorem2}}\label{app:alternative}

We shall show that, for any binary-input channel $p_{Y|X,S}$, state distribution $p_S$, and side channel $p_{\tilde{S}|S}$,
\begin{align*}
C'(p_{Y|X,S},p_S,p_{\tilde{S}|S})=\overline{C}(p_{Y|X,S},p_S)
\end{align*}
if
\begin{align}
H(S|\tilde{S})\leq\frac{4\rho\log 2}{3+2(e-1)\sqrt{2|\mathcal{S}|}}.\label{eq:alternative}
\end{align}

\begin{lemma}\label{lem:newthreshold}
$p_{\hat{X}|S}$  is a stochastically degraded version of $p_{\tilde{S}|S}$ if
\begin{align}
H(S|\tilde{S})\leq\frac{4\tau\rho\log2}{3\tau+2\sqrt{2|\mathcal{S}|}}, \label{eq:anotherbound}
\end{align}
where
\begin{align*}
\tau=\min\limits_{x\in\mathcal{X}_+}\frac{\min_{s\in\mathcal{S}}p_{\hat{X}|S}(x|s)}{\max_{s\in\mathcal{S}}p_{\hat{X}|S}(x|s)}.
\end{align*}
\end{lemma}
\begin{IEEEproof}
Let $\hat{S}$ denote the maximum likelihood estimate of $S$ based on $\tilde{S}$. It suffices to show that $p_{\hat{S}|S}$ is invertible and $p^{-1}_{\hat{S}|S}p_{\hat{X}|S}$ is a valid probability transition matrix if (\ref{eq:anotherbound}) is satisfied.

\begin{table*}[!htbp]
	\caption{Specification of  $\psi(\cdot,\cdot)$ for $\mathcal{U}=\{0,1,\cdots,7\}$ and $\tilde{\mathcal{S}}=\{0,1,\ast\}$}
	\label{tab1}
	\centering
\begin{tabular}{|c|c|c|c|}
  \hline
  $\psi(u,\tilde{s})$ & $\tilde{s}=0$ & $\tilde{s}=1$ & $\tilde{s}=\ast$ \\
\hline
  $u=0$ & 0 & 0 & 0 \\
\hline
  $u=1$ & 1 & 1 & 1 \\
\hline
 $u=2$ & 1 & 1 & 0 \\
\hline
 $u=3$ & 0 & 0 & 1 \\
\hline
  $u=4$ & 0 & 1 & 0 \\
  \hline
 $u=5$ & 0 & 1 & 1 \\
\hline
 $u=6$ & 1 & 0 & 0 \\
\hline
 $u=7$ & 1 & 0 & 1 \\
\hline
\end{tabular}
\end{table*}

Let $\sigma_{\min}(p_{\hat{S}|S})$ denote the smallest singular value of $p_{\hat{S}|S}$. It follows from \cite[Th. 3]{Johnson89} that
\begin{align}
&\sigma_{\min}(p_{\hat{S}|S})\geq\min\limits_{s\in\mathcal{S}}\frac{1}{2}\Bigg(2p_{\hat{S}|S}(s|s)-\sum\limits_{\hat{s}\in\mathcal{S}:\hat{s}\neq s}p_{\hat{S}|S}(\hat{s}|s)\nonumber\\
&\hspace{1.7in}-\sum\limits_{\hat{s}\in\mathcal{S}:\hat{s}\neq s}p_{\hat{S}|S}(s|\hat{s})\Bigg).\label{eq:singularvalue}
\end{align}
Clearly,
\begin{align}
&\min\limits_{s\in\mathcal{S}}\frac{1}{2}\Bigg(2p_{\hat{S}|S}(s|s)-\sum\limits_{\hat{s}\in\mathcal{S}:\hat{s}\neq s}p_{\hat{S}|S}(\hat{s}|s)-\sum\limits_{\hat{s}\in\mathcal{S}:\hat{s}\neq s}p_{\hat{S}|S}(s|\hat{s})\Bigg)\nonumber\\
&=\min\limits_{s\in\mathcal{S}}\frac{1}{2}\Bigg(2-3\sum\limits_{\hat{s}\in\mathcal{S}:\hat{s}\neq s}p_{\hat{S}|S}(\hat{s}|s)-\sum\limits_{\hat{s}\in\mathcal{S}:\hat{s}\neq s}p_{\hat{S}|S}(s|\hat{s})\Bigg)\nonumber\\
&\geq 1-\frac{3}{2}\sum\limits_{s,\hat{s}\in\mathcal{S}:s\neq\hat{s}}p_{\hat{S}|S}(\hat{s}|s).\label{eq:offdiag}
\end{align}
Substituting (\ref{eq:offdiag}) into (\ref{eq:singularvalue}) and invoking (\ref{eq:tbinvokenew})
gives
\begin{align}
\sigma_{\min}(p_{\hat{S}|S})\geq 1-\frac{3H(S|\tilde{S})}{4\rho\log 2}.\label{eq:singlower}
\end{align}
Therefore, $p_{\hat{S}|S}$ is invertible if $H(S|\tilde{S})<\frac{4\rho\log 2}{3}$. Let  $\|\cdot\|_{\infty}$, $\|\cdot\|_2$, and $\|\cdot\|_F$ denote  the maximum row sum matrix norm, the spectral norm, and the Frobenius norm, respectively \cite{HJ85}. Note that
\begin{align}
&\|p^{-1}_{\hat{S}|S}-\mbox{diag}(1,\cdots,1)\|_{\infty}\nonumber\\
&\leq\sqrt{|\mathcal{S}|}\|p^{-1}_{\hat{S}|S}-\mbox{diag}(1,\cdots,1)\|_2\nonumber\\
&\leq\sqrt{|\mathcal{S}|}\|p^{-1}_{\hat{S}|S}\|_2\|p_{\hat{S}|S}-\mbox{diag}(1,\cdots,1)\|_2\label{eq:submultiplicative}\\
&\leq\sqrt{|\mathcal{S}|}\|p^{-1}_{\hat{S}|S}\|_2\|p_{\hat{S}|S}-\mbox{diag}(1,\cdots,1)\|_F,\label{eq:normproduct}
\end{align}
where (\ref{eq:submultiplicative}) follows by the sub-multiplicative property of the spectral norm. We have
\begin{align}
\|p^{-1}_{\hat{S}|S}\|_2&=\frac{1}{\sigma_{\min}(p_{\hat{S}|S})}\nonumber\\
&\leq\Bigg(1-\frac{3H(S|\tilde{S})}{4\rho\log 2}\Bigg)^{-1},\label{eq:duetosinglower}
\end{align}
where (\ref{eq:duetosinglower}) is due to (\ref{eq:singlower}). For $p_{\hat{S}|S}-\mbox{diag}(1,\cdots,1)$, it is clear that the diagonal entries are non-positive, the off-diagonal entries are non-negative, and the sum of all entries is equal to 0; moreover, the sum of its off-diagonal entries is bounded above by $\frac{H(S|\tilde{S})}{2\rho\log 2}$ (see (\ref{eq:tbinvokenew})). Therefore,
\begin{align}
&\|p_{\hat{S}|S}-\mbox{diag}(1,\cdots,1)\|_F\nonumber\\
&=\sqrt{\sum\limits_{s\in\mathcal{S}}(p_{\hat{S}|S}(s|s)-1)^2+\sum\limits_{s,\hat{s}\in\mathcal{S}:s\neq\hat{s}}(p_{\hat{S}|S}(\hat{s}|s))^2}\nonumber\\
&\leq\sqrt{\Bigg(\sum\limits_{s\in\mathcal{S}}(p_{\hat{S}|S}(s|s)-1)\Bigg)^2+\Bigg(\sum\limits_{s,\hat{s}\in\mathcal{S}:s\neq\hat{s}}p_{\hat{S}|S}(\hat{s}|s)\Bigg)^2}\nonumber\\
&=\sqrt{2\Bigg(\sum\limits_{s,\hat{s}\in\mathcal{S}:s\neq\hat{s}}p_{\hat{S}|S}(\hat{s}|s)\Bigg)^2}\nonumber\\
&\leq\frac{H(S|\tilde{S})}{\sqrt{2}\rho\log 2}.\label{eq:Fnupper}
\end{align}
Substituting (\ref{eq:duetosinglower}) and (\ref{eq:Fnupper}) into (\ref{eq:normproduct}) yields
\begin{align}
&\|p^{-1}_{\hat{S}|S}-\mbox{diag}(1,\cdots,1)\|_{\infty}\nonumber\\
&\leq\frac{\sqrt{|\mathcal{S}|}H(S|\tilde{S})}{\sqrt{2}\rho\log 2}\Bigg(1-\frac{3H(S|\tilde{S})}{4\rho\log 2}\Bigg)^{-1}.\label{eq:infnormbound}
\end{align}
To ensure that all entries of $p^{-1}_{\hat{S}|S}p_{\hat{X}|S}$ are non-negative (or equivalently $(\mbox{diag}(1,\cdots,1)-p^{-1}_{\hat{S}|S})p_{\hat{X}|S}$ is component-wise dominated by $p_{\hat{X}|S}$), it suffices to have
\begin{align}
\|p^{-1}_{\hat{S}|S}-\mbox{diag}(1,\cdots,1)\|_{\infty}\leq\tau.\label{eq:posconstraint}
\end{align}
Combining (\ref{eq:infnormbound}) and (\ref{eq:posconstraint}) shows that $p^{-1}_{\hat{S}|S}p_{\hat{X}|S}$ is a valid probability transition matrix\footnote{The requirement that the entries in each row of $p^{-1}_{\hat{S}|S}p_{\hat{X}|S}$ add up to 1 is automatically satisfied.} if  (\ref{eq:anotherbound}) is satisfied\footnote{Note that (\ref{eq:anotherbound}) implies $H(S|\tilde{S})<\frac{4\rho\log 2}{3}$, which further implies the existence of $p^{-1}_{\hat{S}|S}$.}.
\end{IEEEproof}

Since $|\mathcal{X}|=2$, it follows from \cite[Th. 2]{SF04} that there exists $p_{\hat{X}|S}\in\mathcal{P}$ satisfying
\begin{align*}
p_{\hat{X}|S}(x|s)>e^{-1},\quad x\in\mathcal{X}, s\in\mathcal{S}.
\end{align*}
For such $p_{\hat{X}|S}$, we have
\begin{align*}
\tau\geq\frac{1}{e-1}.
\end{align*}
Invoking Lemma \ref{lem:newthreshold} shows that $p_{\hat{X}|S}$  is a stochastically degraded version of $p_{\tilde{S}|S}$ (and consequently $C'(p_{Y|X,S},p_S,p_{\tilde{S}|S})=\overline{C}(p_{Y|X,S},p_S)$) if (\ref{eq:alternative}) is satisfied.

\section{Proof of (\ref{eq:proofinAppB1}) and (\ref{eq:proofinAppB2})}\label{app:underline}

\begin{table*}[!htbp]
	\caption{Specification of  $\psi(\cdot,\cdot)$ for $\mathcal{U}=\{0,1,\cdots,3\}$ and $\tilde{\mathcal{S}}=\{0,1,\ast\}$}
	\label{tab2}
	\centering
\begin{tabular}{|c|c|c|c|}
  \hline
  $\psi(u,\tilde{s})$ & $\tilde{s}=0$ & $\tilde{s}=1$ \\
\hline
  $u=0$ & 0 & 0  \\
\hline
  $u=1$ & 1 & 1  \\
\hline
  $u=2$ & 0 & 1 \\
\hline
  $u=3$ & 1 & 0  \\
  \hline
\end{tabular}
\end{table*}

\begin{lemma}\label{lem:aninequality}
For $\theta\in(0,1)$,
\begin{align*}
\eta(\theta)\triangleq(1-\theta)\log(1+\theta)+\theta\log\theta<0.
\end{align*}
\end{lemma}
\begin{IEEEproof}
We have
\begin{align*}
\frac{\mathrm{d}^2\eta(\theta)}{\mathrm{d}\theta^2}&=\frac{\mathrm{d}}{\mathrm{d}\theta}\Bigg(-\log(1+\theta)+\frac{1-\theta}{1+\theta}+\log\theta+1\Bigg)\\
&=-\frac{1}{1+\theta}-\frac{2}{(1+\theta)^2}+\frac{1}{\theta}\\
&=\frac{1-\theta}{\theta(1+\theta)^2}\\
&>0,\quad\theta\in(0,1),
\end{align*}
which, together with the fact $\eta(0)=\eta(1)=0$, implies the desired result.
\end{IEEEproof}

When $\theta=0$ or $\theta=1$, we have $\underline{C}(p_{Y|X,S},p_S)=\overline{C}(p_{Y|X,S},p_S)$, which implies $\underline{\epsilon}(p_{Y|X,S},p_S)=\underline{q}(p_{Y|X,S},p_S)=0$. When $\theta\in(0,1)$, the maximizer of the optimization problem in (\ref{eq:nostate}), denoted by $p_{\hat{X}}$, is unique and is given by
\begin{align*}
p_{\hat{X}}(0)=p_{\hat{X}}(1)=\frac{1}{2}.
\end{align*}
Now consider $\psi(\cdot,\cdot)$ specified by Table~\ref{tab1}.
It can be verified that
\begin{align*}
&D_{GE}(p_{\hat{U}},\epsilon,u)\\
&=\frac{1}{2}\Bigg((1-\theta)\log2+\log\frac{2}{1+\theta}+\theta\log\frac{2\theta}{1+\theta}\Bigg),\quad u=0,1,\\
&D_{GE}(p_{\hat{U}},\epsilon,u)\\
&=\frac{1}{2}\Bigg(\epsilon(1-\theta)\log2\epsilon+(\theta+\epsilon(1-\theta))\log\frac{2(\theta+\epsilon(1-\theta))}{1+\theta}\\
&\quad+(1-\epsilon(1-\theta))\log\frac{2(1-\epsilon(1-\theta))}{1+\theta}\\
&\quad+(1-\epsilon)(1-\theta)\log2(1-\epsilon)\Bigg),\quad u=2,3,\\
&D_{GE}(p_{\hat{U}},\epsilon,u)\\
&=\frac{1}{2}\Bigg((1-\theta)\log2+(\theta+\epsilon(1-\theta))\log\frac{2(\theta+\epsilon(1-\theta))}{1+\theta}\\
&\quad+\theta\log\frac{2\theta}{1+\theta}+(1-\epsilon)(1-\theta)\log2(1-\epsilon)\Bigg),\quad u=4,5,\\
&D_{GE}(p_{\hat{U}},\epsilon,u)\\
&=\frac{1}{2}\Bigg(\epsilon(1-\theta)\log2\epsilon+\log\frac{2}{1+\theta}\\
&\quad+(1-\epsilon(1-\theta))\log\frac{2(1-\epsilon(1-\theta))}{1+\theta}\Bigg),\quad u=6,7.
\end{align*}
Moreover,
\begin{align}
D_{GE}(p_{\hat{U}},0,u)&=\frac{1}{2}\Bigg((1-\theta)\log2+\log\frac{2}{1+\theta}\nonumber\\
&\hspace{1.0in}+\theta\log\frac{2\theta}{1+\theta}\Bigg)\nonumber\\
&=\underline{C}(p_{Y|X,S},p_S),\quad u=0,1,2,3,\nonumber\\
D_{GE}(p_{\hat{U}},0,u)&=(1-\theta)\log2+\theta\log\frac{2\theta}{1+\theta}\nonumber\\
&<\underline{C}(p_{Y|X,S},p_S),\quad u=4,5,\label{eq:invokeineq1}\\
D_{GE}(p_{\hat{U}},0,u)&=\log\frac{2}{1+\theta}\nonumber\\
&>\underline{C}(p_{Y|X,S},p_S),\quad u=6,7,\label{eq:invokeineq2}
\end{align}
where (\ref{eq:invokeineq1}) and (\ref{eq:invokeineq2}) follow from Lemma \ref{lem:aninequality}.
Therefore, we have
\begin{align*}
&\epsilon(u)=0,\quad u=0,1,2,3,4,5,\\
&\epsilon(u)=\hat{\epsilon}(\theta),\quad u=6,7,
\end{align*}
which, together with (\ref{eq:maxepsilon}), proves (\ref{eq:proofinAppB1}) for $\theta\in(0,1)$.
Next consider $\psi(\cdot,\cdot)$ specified by Table~\ref{tab2}. It can be verified that
\begin{align*}
&D_{GS}(p_{\hat{U}},q,u)\\
&=\frac{1}{2}\Bigg((1-\theta)\log2+\log\frac{2}{1+\theta}+\theta\log\frac{2\theta}{1+\theta}\Bigg),\quad u=0,1,\\
&D_{GS}(p_{\hat{U}},q,2)\\
&=(1-q)(1-\theta)\log2(1-q)\\
&\quad+(\theta+q(1-\theta))\log\frac{2(\theta+q(1-\theta))}{1+\theta},\\
&D_{GS}(p_{\hat{U}},q,3)\\
&=q(1-\theta)\log2q+(1-q(1-\theta))\log\frac{2(1-q(1-\theta))}{1+\theta}.
\end{align*}
Moreover,
\begin{align}
D_{GS}(p_{\hat{U}},0,u)&=\frac{1}{2}\Bigg((1-\theta)\log2+\log\frac{2}{1+\theta}\nonumber\\
&\hspace{1in}+\theta\log\frac{2\theta}{1+\theta}\Bigg)\nonumber\\
&=\underline{C}(p_{Y|X,S},p_S),\quad u=0,1,\nonumber\\
D_{GS}(p_{\hat{U}},0,2)&=(1-\theta)\log2+\theta\log\frac{2\theta}{1+\theta}\nonumber\\
&<\underline{C}(p_{Y|X,S},p_S),\label{eq:invokeineq1again}\\
D_{GS}(p_{\hat{U}},0,3)&=\log\frac{2}{1+\theta}\nonumber\\
&>\underline{C}(p_{Y|X,S},p_S),\label{eq:invokeineq2again}
\end{align}
where (\ref{eq:invokeineq1again}) and (\ref{eq:invokeineq2again}) follow from Lemma \ref{lem:aninequality}. Therefore, we have
\begin{align*}
&q(u)=0,\quad u=0,1,2,\\
&q(3)=\hat{q}(\theta),
\end{align*}
which, together with (\ref{eq:maxq}), proves (\ref{eq:proofinAppB2}) for $\theta\in(0,1)$.


\section{Proof of (\ref{eq:proofinAppC1}) and (\ref{eq:proofinAppC2})}\label{app:overline}

When $\theta=0$ or $\theta=1$, we have $\underline{C}(p_{Y|X,S},p_S)=\overline{C}(p_{Y|X,S},p_S)$, which implies $\overline{\epsilon}(p_{Y|X,S},p_S)=1$ and $\overline{q}(p_{Y|X,S},p_S)=\frac{1}{2}$.
When $\theta\in(0,1)$,  the maximizer of the optimization problem in (\ref{eq:perfectstate}), denoted by $p_{\hat{X}|S}$, is unique and is given by
\begin{align*}
&p_{\hat{X}|S}(x|s)\\
&=\left\{
                     \begin{array}{ll}
                       \Bigg(1+(1-\theta)\theta^{\frac{\theta}{1-\theta}}\Bigg)^{-1}\theta^{\frac{\theta}{1-\theta}}, & x=s, \\
                       \Bigg(1+(1-\theta)\theta^{\frac{\theta}{1-\theta}}\Bigg)^{-1}\Bigg(1-\theta^{\frac{1}{1-\theta}}\Bigg), & \mbox{otherwise}.
                     \end{array}
                   \right.
\end{align*}
In view of (\ref{eq:invewepsilon}) and (\ref{eq:inviewq}), it suffices to show that
\begin{align*}
\theta^{\frac{\theta}{1-\theta}}<1-\theta^{\frac{1}{1-\theta}},\quad\theta\in(0,1).
\end{align*}
Indeed, for $\theta\in(0,1)$,
\begin{align*}
&\theta^{\frac{\theta}{1-\theta}}<1-\theta^{\frac{1}{1-\theta}}\\
&\Leftrightarrow 1<\theta^{-\frac{\theta}{1-\theta}}-\theta\\
&\Leftrightarrow(1-\theta)\log(1+\theta)+\theta\log\theta<0,
\end{align*}
and the last inequality is true according to Lemma \ref{lem:aninequality}.

\section*{Acknowledgment}

The authors wish to thank the associate editor and the anonymous reviewer for their valuable comments and suggestions.

\end{document}